\documentclass[pra,twocolumn,english,showkeys,superscriptaddress,floatfix,longbibliography,aps,amsmath,amssymb]{revtex4-2}

\usepackage[english]{babel}
\usepackage[utf8]{inputenc}
\usepackage{soul}
\usepackage{amsthm}
\usepackage{mathtools}
\usepackage{physics}
\usepackage{xcolor}
\usepackage{graphicx}
\usepackage[left=23mm,right=13mm,top=35mm,columnsep=15pt]{geometry} 
\usepackage{adjustbox}
\usepackage{placeins}
\usepackage[T1]{fontenc}
\usepackage{lipsum}
\usepackage{csquotes}
\usepackage[pdftex, pdftitle={Article}, pdfauthor={Author}]{hyperref} 
\usepackage{simpler-wick}
\usepackage{listings}
\usepackage{nccmath}
\usepackage{comment}
\usepackage{multirow}
\usepackage{relsize}
\usepackage{wasysym}
\usepackage{natbib}

\definecolor{codegreen}{rgb}{0,0.6,0}
\definecolor{codegray}{rgb}{0.5,0.5,0.5}
\definecolor{codepurple}{rgb}{0.58,0,0.82}
\definecolor{backcolour}{rgb}
{0.95,0.95,0.92}
\definecolor{ForestGreen}{RGB}{34,139,34}

\lstdefinestyle{mystyle}{
    backgroundcolor=\color{backcolour},   
    commentstyle=\color{codegreen},
    keywordstyle=\color{magenta},
    numberstyle=\tiny\color{codegray},
    stringstyle=\color{codepurple},
    basicstyle=\ttfamily\footnotesize,
    breakatwhitespace=false,         
    breaklines=true,                 
    captionpos=b,                    
    keepspaces=true,                 
    numbers=left,                    
    numbersep=5pt,                  
    showspaces=false,                
    showstringspaces=false,
    showtabs=false,                  
    tabsize=2
}
\lstdefinelanguage{Julia}%
  {morekeywords={abstract,break,case,catch,const,continue,do,else,elseif,%
      end,export,false,for,function,immutable,import,importall,if,in,%
      macro,module,otherwise,quote,return,switch,true,try,type,typealias,%
      using,while},%
   sensitive=true,%
   alsoother={},%
   morecomment=[l]\#,%
   morecomment=[n]{\#=}{=\#},%
   morestring=[s]{"}{"},%
   morestring=[m]{'}{'},%
}[keywords,comments,strings]%

\usepackage[utf8]{inputenc}
\usepackage{xcolor}
\usepackage{enumitem}
\usepackage{booktabs}
\usepackage{caption}

\begin{document}
\title{High-fidelity Quantum Readout Processing via an Embedded SNAIL Amplifier}
\author{Leon Bello}
\email[Correspondence email address: ]{lbello@princeton.edu}
\affiliation{Department of Electrical and Computer Engineering, Princeton University}
\author{Boris Mesits}
\affiliation{Department of Physics and Astronomy, University of Pittsburgh}
\affiliation{Department of Applied Physics, Yale University}
\author{Michael Hatridge}
\affiliation{Department of Applied Physics, Yale University}
\author{Hakan E. Türeci}
\affiliation{Department of Electrical and Computer Engineering, Princeton University}
    
\date{\today} 

\newcommand{\boris}[1]{\textcolor{red}{#1}}

\begin{abstract}
Scalable, high-fidelity quantum-state readout remains a central challenge in the development of large-scale superconducting quantum processors. Conventional dispersive readout architectures depend on bulky isolators and external amplifiers, introducing significant hardware overhead and limiting opportunities for on-chip information processing. In this work, we propose a novel approach that embeds a nonlinear Superconducting Nonlinear Asymmetric Inductive eLement (SNAIL) into the readout chain, enabling coherent and directional processing of readout signals directly on-chip. This embedded SNAIL platform allows frequency-multiplexed resonators to interact through engineered couplings, forming a tunable readout-amplifier-output architecture that can manipulate quantum readout data \textit{in situ}. Through theoretical modeling and numerical optimization, we show that this platform enhances fidelity, suppresses measurement-induced decoherence, and simplifies hardware complexity. These results establish the hybridized SNAIL as a promising building block for scalable and coherent quantum-state readout in next-generation processors.
\end{abstract}


\maketitle
\label{sec:intro} 
\section{Introduction}
Readout of superconducting circuits relies on a dispersive interaction between a qubit and a readout resonator. The qubit state is then inferred from the state-dependent frequency shift of the microwave resonator. This technique is central to circuit QED and has been widely adopted in present-day superconducting quantum processors due to its straightforward implementation and compatibility with existing hardware platforms ~\cite{blais_cavity_2004}. 

The standard approach, which we denote as \emph{Conventional Dispersive Readout} (CDR), infers the qubit state from the state-dependent frequency shift of a microwave resonator coupled dispersively to the qubit. While this measurement modality is central to circuit QED and has been widely adopted in present-day superconducting quantum processors due to its straightforward implementation~\cite{blais_circuit_2021,clerk_introduction_2010}.

First, the hardware overhead becomes prohibitive. In the conventional chain, the
readout resonator is coupled to the first amplifier via a cascade of off-chip
non-reciprocal components, typically ferrite circulators and isolators \cite{pozar_microwave_2012}. The requisite isolators are bulky, consuming valuable cryogenic volume, and introduce significant thermal load and flux noise close to the qubit plane. 

Second, the measurement fidelity is fundamentally limited by the quantum efficiency of the detection chain and the readout architecture itself. Signal attenuation from intervening isolators and the added noise of standard amplifiers degrade the signal-to-noise ratio (SNR)~\cite{friis_noise_1944, caves_quantum_1982}. Furthermore, in CDR, the readout resonator is forced to play a dual role: it serves as the mediator of information to the outside world, but also as the stage on which this information is accumulated. This conflation couples the signal extraction rate to the information storage time, preventing independent optimization of the readout dynamics \cite{clerk_introduction_2010}.

To address these limitations, recent efforts have explored replacing static, bulky
components with tunable, on-chip alternatives. Quantum-limited amplification is now routinely achieved using traveling-wave or resonator-based parametric amplifiers that offer $\sim$20\,dB gain with minimal added noise~\cite{macklin_nearquantum-limited_2015}. However, standard amplifiers are non-directional and require isolation to prevent back-action. Consequently, a promising strategy under active development is the integration of \textit{on-chip directional amplifiers}. Several architectures have been demonstrated, including parametric conversion with engineered dissipation~\cite{metelmann_quantum-limited_2022}, engineered stopbands in traveling-wave amplifiers~\cite{malnou_traveling-wave_2024}, and wave-interference schemes~\cite{lecocq_nonreciprocal_2017,sliwa_reconfigurable_2015}. These implementations typically fall into the class of \textit{steady-state methods}, where gain and conversion pumps remain on throughout the readout and directionality arises from driven-dissipative interactions. While effective, such approaches remain constrained by the complexity of the required pump schemes and their reliance on timescale separation. 

Crucially, these innovations have largely preserved the "component-based" paradigm, treating the readout resonator, the directional element, and the amplifier as discrete, cascaded modules. In this work, we propose a rethinking of the measurement chain. Rather than cascading discrete components, we introduce an architecture where the readout resonator \emph{itself} embeds the non-linearity required for amplification and directionality.

\begin{figure}[!h]
    \centering
    \includegraphics[width=\linewidth]{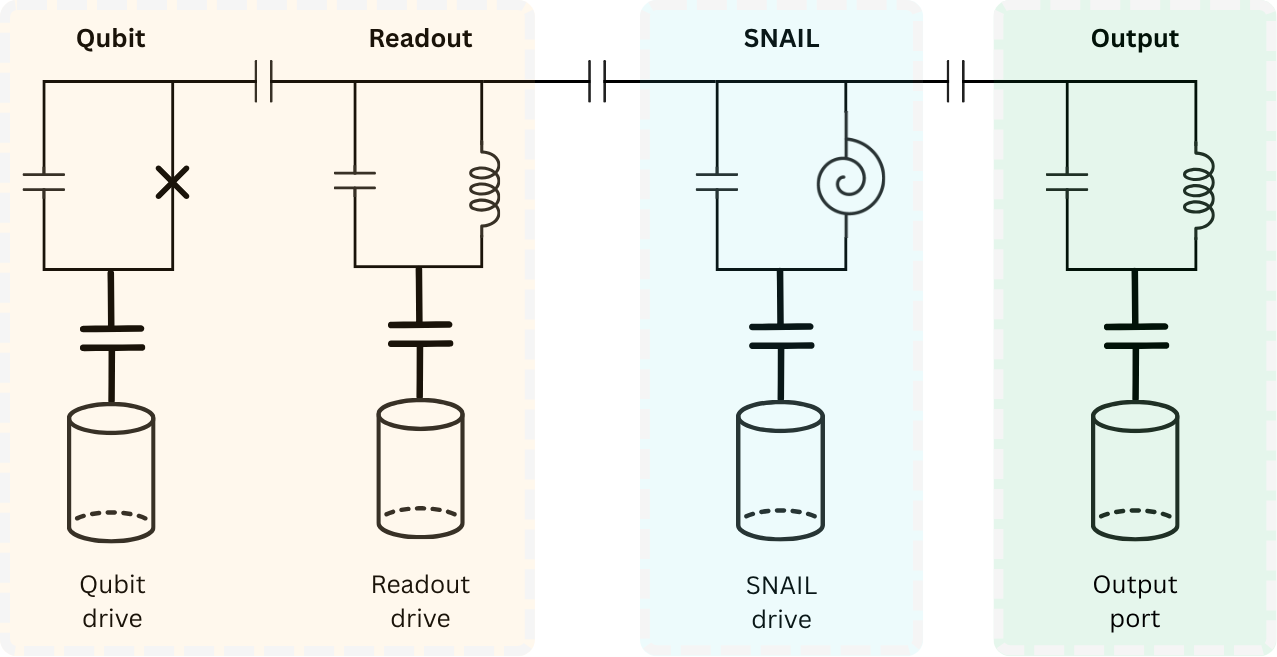}
    \caption{A circuit schematic of the embedded amplifiers platform. A qubit-resonator system is coupled to a nonlinear SNAIL resonator, which coupled to a fast output resonator. The nonlinearity of the SNAIL enables frequency multiplexed interactions between the resonators.}
    \label{fig:schematic}
\end{figure}

Our proposed architecture embeds a Superconducting Nonlinear Asymmetric Inductive eLement (SNAIL) amplifier on-chip, coupled off-resonance to both a readout resonator and an output resonator, as illustrated in Fig.~\ref{fig:schematic}. This weak hybridization enables tunable, \textit{pulsed frequency-multiplexed parametric interactions}, allowing for \textit{unitary manipulation} of the readout field itself with effectively no losses. This platform enables a "Catch-Process-Release" protocol that temporally decouples information accumulation from signal extraction. By processing the readout signal \textit{in situ}—performing directional amplification, selective noise shaping, and programmable conversion—before it enters the lossy transmission line, our architecture maximizes fidelity and reduces hardware complexity. This effectively transforms the readout chain from a passive signal extractor to an active information-processing layer
integrated within the quantum hardware. We note that a similar approach was explored by Rosenthal et al.~\cite{rosenthal_efficient_2021}, albeit in a different design and with a different focus~\cite{gard_fast_2024}.

It is important to distinguish this approach from similar parametric architectures used in
modular quantum computing, such as the "Quantum State Router"~\cite{zhou_realizing_2023}.
While those systems also employ SNAIL-mediated three-wave mixing, their primary objective
is coherent state transfer (e.g., SWAP gates) between high-Q communication modes to link
distinct quantum modules. In contrast, our work adapts this nonlinearity specifically for the \emph{readout} chain. Rather than routing quantum information between modules, we utilize the SNAIL to actively process the readout signal \emph{in situ}—generating phase-sensitive gain and squeezing—thereby transforming the readout resonator from a passive storage element into an active, directional amplification stage.

This paper is organized as follows: Sec.~\ref{sec:EA_platform} introduces our architecture and theoretical model. Sec.~\ref{sec:sequential_scheme} presents an optimized pulse sequence for coherent readout processing. Sec. \ref{sec:output_fidelity}  analyzes the readout fidelity achievable using our method and the achievable advantage compared to CDR. Sec.~\ref{sec:non-idealities} addresses non-idealities and scalability. We conclude in Sec.~\ref{sec:conclusions} with prospects for future applications.

\section{Device architecture and modeling}
\label{sec:EA_platform}
\subsection{Effective Hamiltonian model}
We propose a modular architecture for scalable, frequency-multiplexed quantum-state readout based on a common parametric mixing element. While this topology shares the SNAIL-mediated couplings of 'quantum state routers' \cite{zhou_realizing_2023}, we repurpose the nonlinearity for active readout processing—specifically gain and squeezing—rather than coherent state transfer between computing modules.

The core of our device is a superconducting nonlinear asymmetric inductive element (SNAIL), a three-wave mixing device which mediates interactions among multiple resonators on-chip. Each readout unit comprises three detuned but weakly coupled modes: (1) a high-Q readout resonator (\(\hat{a}\)) dispersively coupled to a qubit, (2) a low-Q output resonator (\(\hat{b}\)) for fast signal extraction, and (3) a nonlinear SNAIL-hosting mode that acts as a tunable mixing element. The detuning between modes suppresses unwanted crosstalk, while the weak hybridization with the SNAIL preserves the nonlinearity across the system. 

More precisely, the Hamiltonian of the device in the lab frame can be written as,
\begin{subequations}
\begin{align}
    &\hat{H}_0 = (\omega_a \pm \chi) \hat{a}^\dagger \hat{a} + \omega_b \hat{b}^\dagger \hat{b} + \omega_s \hat{s}^\dagger \hat{s}, \\
    &\hat{H}_L = g_{as} \hat{a}^\dagger \hat{s} + g_{bs} \hat{b}^\dagger \hat{s} + g_{ab} \hat{a}^\dagger \hat{b} + h.c., \\
    &\hat{H}_{\text{NL}} = g_{sss}(\hat{s} + \hat{s}^\dagger)^3, \\
    &\hat{H}_{\text{drive}} = \eta(t)^*\hat{a}  + p^*(t) \hat{s} + h.c.
\end{align}
\end{subequations}
where $\hat{a}$, $\hat{b}$, and $\hat{s}$ represent the mode operators for the readout, output, and SNAIL resonators, respectively. The readout resonator is dispersively coupled to a qubit, shifting its resonance frequency by an amount $\chi$ depending on the qubit state.
\begin{table}[h!]
\captionsetup{font=small}
\small
\centering
\begin{tabular}{@{}lcc@{}}
\toprule
\textbf{Mode} & $\omega_i/2\pi$ [GHz] & \quad $\gamma_i$ [MHz] \\ \midrule
Readout & 6.0 & 0.1 \\
SNAIL & 4.0 & 1.0 \\
Output & 7.5 & 20.0 \\
\bottomrule
\end{tabular}
\caption{Resonant frequencies and decay rates for different amplifier components. The dispersive shift is assumed to be around $\chi = 3\text{MHz}$. The values listed are representative parameters relevant to an experimental realization, chosen to benchmark the theoretical model presented here.}
\label{tab:sys_parameters}
\end{table}
\begin{figure}
    \centering
    \includegraphics[width=\linewidth]{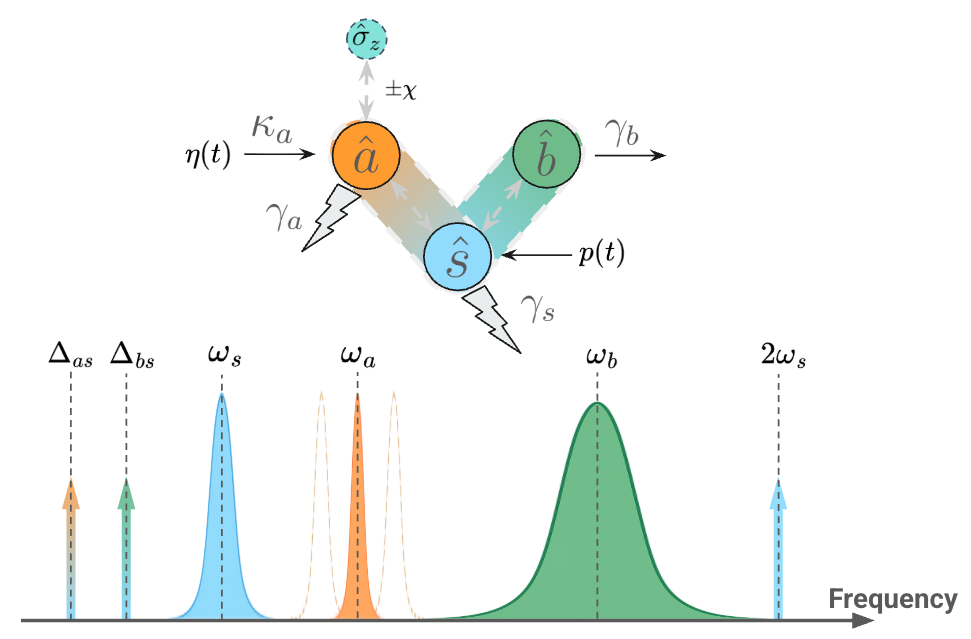}
    \caption{\textbf{(top)} A mode connectivity schematic and physical couplings (denoted by the grey arrows).  \textbf{(bottom)} Spectrum of the modes. The Lorentzians correspond to the resonator spectral responses, with colors matching their corresponding modes. The colored arrows represent pumps corresponding to different parametric processes, with $\Delta_{ij} = |\omega_i - \omega_j|$.}
    \label{fig:mode-schematic}
\end{figure}

We assume operation in the linear regime, where pump depletion is negligible, and rotating-wave approximation (RWA) is justified. For the numerical simulations in this theoretical study, we adopt the system parameters listed in  \ref{tab:sys_parameters}; these are representative values relevant to a typical experimental realization. 

Transforming into the rotating frame of the bare resonances, diagonalizing the linear Hamiltonian, and displacing the pump mode, we obtain the effective quadratic Hamiltonian (see App.~\ref{app:effective_Hamiltonian_derivation} for details). These quadratic interactions can induce either conversion or amplification:
\begin{subequations}
    \begin{equation}
        \hat{H}_{\text{amp}} = g_{\omega_p \omega_i \omega_j} \, p_{\omega_p} \hat{a}_{\omega_i} \hat{a}_{\omega_j} e^{-i\Delta_{\omega_p \omega_i \omega_j} t} + \text{h.c.}, 
    \end{equation}
    \begin{equation}
        \hat{H}_{\text{conv}} = g_{\omega_p - \omega_i +\omega_j} \, p_{\omega_p} \hat{a}_{\omega_i}^\dagger \hat{a}_{\omega_j} e^{-i\Delta_{\omega_p \omega_i \omega_j} t} + \text{h.c.}, 
    \end{equation}
\end{subequations}
where we require $ \Delta_{\omega_p\omega_i \omega_j} \equiv \omega_p + \omega_i + \omega_j \ll \min(\omega_p, \omega_i, \omega_j)$.
Here, the subscripts denote the resonator frequencies, e.g., $\hat{a}_{\omega_s} = \hat{s}$, and the detuning is assumed to be much smaller than all time-scales in the system. A comprehensive treatment of measurement-induced backaction is beyond the scope of this work. However, proceeding on the assumption that such effects become increasingly pronounced at high intracavity fields, we specifically restrict our analysis and optimization to the regime of limited photon budgets.

The interaction strengths $g_{\omega_p \omega_i \omega_j}$ depend on the interaction multiplicity $C_{pij}$, the hybridization of the involved modes $\lambda_{\omega_i}$, and the intrinsic SNAIL nonlinearity $g_{sss}$:
\begin{subequations}
\label{eq:effective_couplings}
\begin{equation}
    g_{\omega_p, \omega_i, \omega_j} = C_{pij} \lambda_{\omega_i} \lambda_{\omega_j} g_{sss}  ,
\end{equation}
\begin{equation}
    \lambda_{\omega_i} = \frac{g_{\omega_i \omega_s}}{|\omega_i - \omega_s|}.
\end{equation}
\end{subequations}
The interaction multiplicity $C_{pij}$ is the combinatorial factor arising from the expansion of the cubic Hamiltonian, representing the number of distinct permutations of the field operators involved in the mixing process.
In this definition, the coupling rate refers to the three-wave mixing interaction between the pump and two other modes. Alternatively, the pump amplitude can be absorbed into the coupling rate, yielding an effective, pump-dependent quadratic interaction between the modes $\hat{a}_{\omega_i}$ and $\hat{a}_{\omega_i}$.

By applying microwave drives at carefully chosen frequencies, we selectively activate parametric processes such as mode conversion and amplification. Crucially, the frequency of the applied drives determines which resonator pairs interact, enabling flexible, frequency-multiplexed control over the readout dynamics. By controlling the amplitude and phase of the pump, we can tune the interactions and control the effective coupling rates. This flexible approach allows us to perform very general coherent information processing, and forms the backbone of our architecture, and supports dense integration and scalable readout without requiring bulky off-chip elements, paving the way for embedded, directional amplification architectures.

The generation of tunable interactions via parametric modulation builds upon established techniques in superconducting circuit QED, where frequency-multiplexed pumping has been successfully employed to realize "quantum state routers" for modular interconnects~\cite{zhou_realizing_2023} and reconfigurable directional amplifiers~\cite{sliwa_reconfigurable_2015, lecocq_nonreciprocal_2017}. However, while those architectures primarily utilize three-wave mixing for coherent state transfer or steady-state routing, we adapt this framework to construct a programmable \emph{readout} processor. In this context, the microwave drives do not merely route information; they selectively activate a specific sequence of unitary operations to process the measurement signal \emph{in situ}. This distinction is crucial: rather than linking equivalent quantum modules, our architecture uses the SNAIL as an active interface between the high-Q readout mode and the lossy output, enabling dense integration and directional gain without the hardware overhead of external isolators.

\subsection{Covariance matrix and means description}
Generally, our system is an open system, described by the Lindblad equation,
\begin{equation}
    \frac{\rm d}{\rm dt} \rho = -i[\hat{H}, \rho] + \sum_i \gamma_i \mathcal{D}[\hat{a}_i]
\end{equation}
where $\hat{H}$ is the system Hamiltonian, and $\mathcal{D}[\hat{a}_i]$ is the dissipation super-operator with dissipation rate  $\gamma_i$. 

In the dispersive limit, the system can be modeled as a simple linear (time-varying) system. Crucially, our readout protocol relies on the specific temporal structure of the control fields $\eta(t)$ and $p(t)$. Unlike conventional dispersive readout, where continuous-wave (CW) drives force the system into a steady state, our scheme is predicated on a \emph{pulsed} Hamiltonian interaction $\hat{H}(t)$. By applying discrete sequences of pulses, we engineer the system's \emph{transient} response, selectively activating specific interaction terms (e.g., beam-splitter or squeezing) for fixed durations. This approach allows us to construct the readout as a compiled sequence of unitary gates---catching, processing, and releasing the signal---rather than passive monitoring of a steady-state field.

In that regime, the system admits a simple, low-dimensional description, in terms of the covariance matrix and means of the quadrature amplitudes. Let us define a vector of quadrature operators $\vec{X} = (\hat{q}_r, \hat{p}_r, \hat{q}_s, \hat{p}_s, \hat{q}_o, \hat{p}_o)^\intercal$. As shown in App.~\ref{app:cov_eoms}, since the system is linear, the dynamics can be exactly described by just the evolution of the mean quadrature amplitudes $\vec{\mu}$ and the covariance matrix $\mathbf{V}$,
\begin{align}
    &\vec{\mu} = \langle \vec{X} \rangle =  (\langle \hat{q}_a \rangle,  \langle \hat{p}_a \rangle, \langle \hat{q}_s \rangle, \langle \hat{p}_s \rangle, \langle \hat{q}_b \rangle,  \langle \hat{p}_b \rangle)^\intercal \\
    &[\mathbf{V}]_{ij} = \frac{1}{2} \langle [\hat{X}_i, \hat{X}_j]_+ \rangle - \langle \hat{X}_i \rangle \langle \hat{X}_j \rangle 
\end{align}
where $\hat{q}_i = \frac{1}{\sqrt{2}}(\hat{a}_i + \hat{a}_i^\dagger)$ and $\hat{p}_i = -\frac{i}{\sqrt{2}}(\hat{a}_i - \hat{a}_i^\dagger)$.

Since we are in the linear regime, our Hamiltonian can always be written as a quadratic form,
\begin{equation}
    \hat{H} = \vec{X}^{\intercal} \mathbf{H} \vec{X}
\end{equation}
Additionally, we assume simple resonator decay in each of the different resonators. We arrange the decay rates in a block-diagonal matrix,
\begin{equation}
    \mathbf{\Gamma} = \begin{pmatrix}
    \gamma_a \mathbf{I}_2 & 0_2 & 0_2 \\
    0_2 & \gamma_s \mathbf{I}_2 & 0_2 \\
    0_2 & 0_2 & \gamma_b \mathbf{I}_2
    \end{pmatrix}
\end{equation}

In that case, the equations of motion of the means and covariance are independent, and can be written succinctly,
\begin{subequations}
\label{eq:eoms}
\begin{align}
    &\frac{\rm d}{\rm dt} \vec{\mu} = \left( \Omega \mathbf{H} - \mathbf{\Gamma} \right) \vec{\mu} + \vec{d}(t) \\
    &\frac{\rm d}{\rm dt} \mathbf{V} = \left[ \left( \Omega \mathbf{H} - \mathbf{\Gamma} \right)\mathbf{V} + \mathbf{V}\left( \Omega \mathbf{H} - \mathbf{\Gamma} \right)^\intercal  \right ] + \frac{\mathbf{\Gamma}}{2}
\end{align} 
\end{subequations}

where $\vec{d}(t)$ is a vector of drive amplitudes and $\Omega = \bigoplus_{i=1}^{3} \begin{pmatrix} 0 & 1 \\ -1 & 0\end{pmatrix}$ is the block-diagonal symplectic form matrix, appearing from the commutation relations of the quadrature amplitudes. The equation describing the covariance matrix is the Lyapunov equation, often appearing in the context of stability of dynamical systems. The homogeneous term reflects the "driving" of the covariance matrix by the vacuum fluctuations \cite{chen_linear_1999, wiseman_quantum_2009}.

\subsection{Readout performance metrics}
A key aspect of this platform and work is the optimization of system and pulse parameters. Optimal readout must adhere to different, often conflicting, requirements.

Principally, optimal readout requires maximizing the distinguishability of the
signal distributions. Formally, this is quantified by the Fisher Discriminant.
\begin{equation}
\label{eq:fisher_discriminant}
    \mathcal{D}^2(\vec{\theta}) = \frac{\Delta\vec{\mu}_{eg}^\intercal \Delta\vec{\mu}_{eg}}{\Delta\vec{\mu}_{eg}^\intercal ( \mathbf{V}_e + \mathbf{V}_g ) \Delta\vec{\mu}_{eg}}
\end{equation}
where $\Delta \vec{\mu}_{eg} = \vec{\mu}_e - \vec{\mu}_g$ and the subscript $e,g$ represent the qubit state. The different system and pulse parameters are grouped formally in $\vec\theta$. In the linear regime--where the readout states are Gaussian with equal variances--this metric simplifies to the squared Signal-to-Noise Ratio (SNR). 

We note that the Fisher Discriminant is a linear measure of distinguishability,
relying principally on the separation of the state means relative to their variance. Consequently, it is an insufficient metric for scenarios where the information is encoded primarily in the state covariance (e.g., distinguishing two zero-mean states with different variances). However, in the context of dispersive readout, the interaction explicitly maps the qubit state to a coherent field displacement $\Delta\vec{\mu}$. Since our protocol is designed to maximize this displacement, the readout states remain in the regime where linear separability is the dominant factor, rendering the Fisher Discriminant a robust and efficient objective function. This claim is essentially equivalent to saying that the idealized readout problem can be optimally separated by a linear separator.

\begin{figure}[!h]
    \centering
    \includegraphics[width=\linewidth]{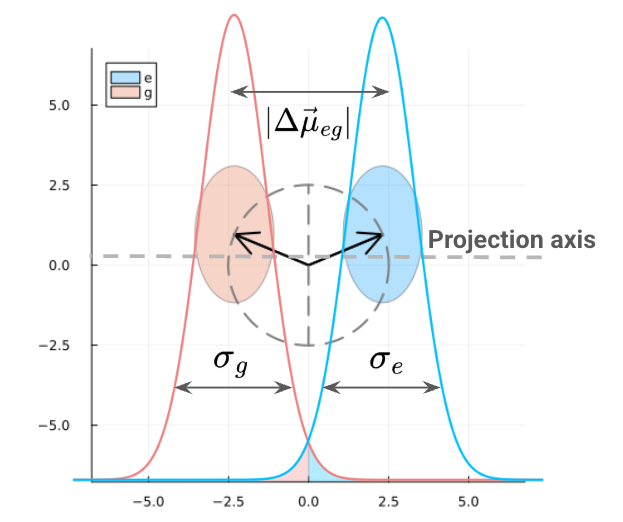}
    \caption{LDA finds an optimal projection axis that maximizes the separation between the two states, in units of the projected noise. The highlighted region represent probability of misclassifications $P_{err}$.}
    \label{fig:fisher_discriminant}
\end{figure}

As illustrated in Figure \ref{fig:fisher_discriminant}, LDA finds an optimal separation axis to distinguish two Gaussian states. The Fisher Discriminant quantifies the separability by comparing the mean separation with projected noise. 

In the case where the projected noises for the two state are the same, this can be related to the more commonly used definition of Fidelity $F = 1 - P_{\rm err}$ by (see App. \ref{app:optimization} for a derivation):
\begin{equation}
\label{eq:error_prob}
    P_{\rm err} =  \rm{erfc} \left( \frac{1}{2} \sqrt{\frac{\mathcal{D}^2(\vec{\theta})}{2}} \right).
\end{equation}
By symmetry this would be the relevant case for our analysis, but we note that in the case where the distributions have different projected variances, the Fisher Discriminant cannot be directly related to the fidelity, as discussed in App. \ref{app:optimization}. 

This relationship highlights the key simplification enabled by the linear regime: since the complementary error function ($\text{erfc}$) in Eq. \ref{eq:error_prob} is strictly monotonic, maximizing the power SNR in Eq. \ref{eq:fisher_discriminant} is mathematically equivalent to maximizing the full readout fidelity $\mathcal{F} = 1 - P_{err}$. Consequently, we adopt the SNR—or equivalently, the Fisher Discriminant—as our primary objective function for optimization, avoiding the computational overhead of minimizing the overlap integral directly.

In dispersive readout, the qubit-dependent rotation of the resonator state simplifies the geometry of state separation; the distributions lie on a circle whose radius depends on the readout resonator occupation, as illustrated in Fig. \ref{fig:fisher_discriminant}. 
This leads to two intuitive criteria for optimal readout in the linear regime: \textbf{(1)} the signal states should be along the radius of the circle, and \textbf{(2)} the noise should be minimized along their separation axis-—ideally through squeezing--so that the projected variance is reduced where it matters most.

While the Fisher Discriminant serves as a generally valid objective function for binary state discrimination, the optimization must be performed within the bounds of physical validity. Unconstrained maximization of the distinguishability could trivially drive the system into regimes where the simplified Hamiltonian description breaks down or where deleterious effects dominate. Consequently, we treat the readout optimization as a \emph{constrained} problem: we maximize the Fisher Discriminant subject to specific physical bounds that ensure the stability and non-destructiveness of the measurement.

First, the intracavity photon number must remain below a critical threshold $n_a^{\rm crit}$ to prevent measurement-induced qubit ionization~\cite{dumas_measurement-induced_2024, khezri_measurement-induced_2023}. Second, the parametric coupling rates are limited to a critical value $g^{\rm crit}_{\omega_1 \omega_2}$ to avoid driving the SNAIL into chaotic instability~\cite{xia_exceeding_2025}. Finally, to ensure the optimized signal-to-noise ratio is preserved against downstream amplifier noise~\cite{friis_noise_1944}, we impose a condition on the minimum signal gain:
\begin{equation}
    \mathcal{G}_{ro}(\vec\theta) =  \sqrt{\frac{\max_t n_s}{\max_tn_a}} \geq \mathcal{G}_{target}
\end{equation}
By enforcing these constraints, we ensure the optimized pulse sequences maximize fidelity while adhering to the "speed limits" and safety margins of the hardware.

\section{Coherent processing of readout information using pulsed interactions}
\label{sec:sequential_scheme}
\begin{figure*}[!t]
    \centering
    \includegraphics[width=0.8\linewidth]{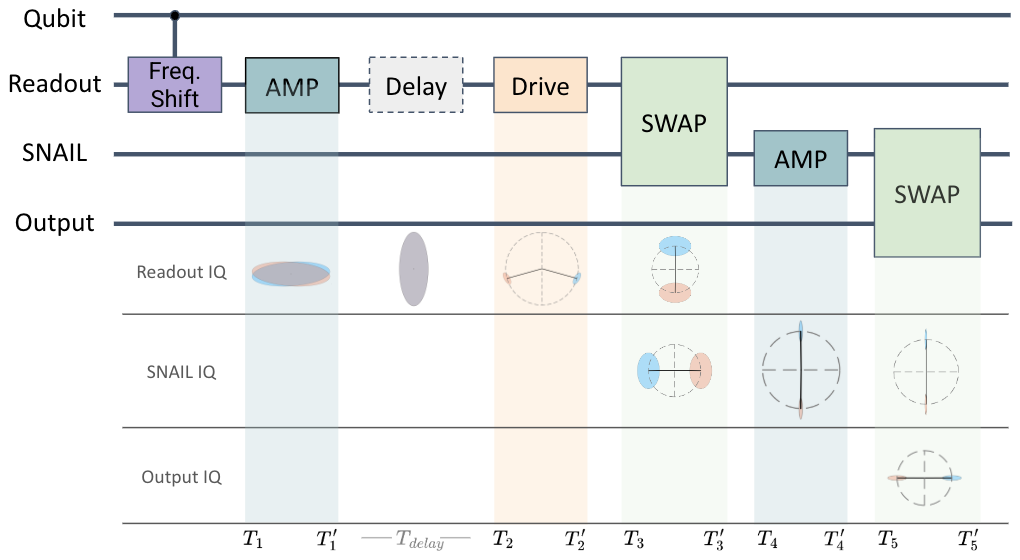}
    \caption{\textbf{Top:} The different parametric interactions can be modeled as a set of bosonic gates, modifying the IQ distributions of the fields. \textbf{Bottom:} The IQ distributions generated by applying the different gates (not to scale). Empty spaces indicate negligible population.}
    \label{fig:seq_scheme}
\end{figure*}
\twocolumngrid
Our parametric interactions enable a complete set of bosonic gates, including mode conversion, phase-sensitive (single-mode) amplification, and phase-preserving (two-mode) amplification. As illustrated in Fig. \ref{fig:seq_scheme}, we can use these gates to control the flow of quantum information between resonators, providing precise manipulation of the readout process.

By carefully timing these interactions, we can coherently transfer the readout information to the output while achieving quantum-limited (or even noiseless) amplification with full directionality. 
This sequential approach divides the readout into discrete stages, each optimized to maximize qubit state discrimination and adhering to the technical constraints discussed above. Directionality is ensured by dynamically controlling the conversion pulse duration to realize a \emph{complete coherent state transfer} (a $\pi$-pulse or full \emph{iSWAP}) -- once the population is fully exchanged to the target resonator, the interactions are switched off, preventing unwanted back-back-conversion.

\begin{figure}[!h]
    \centering
    \includegraphics[width=0.85\linewidth]{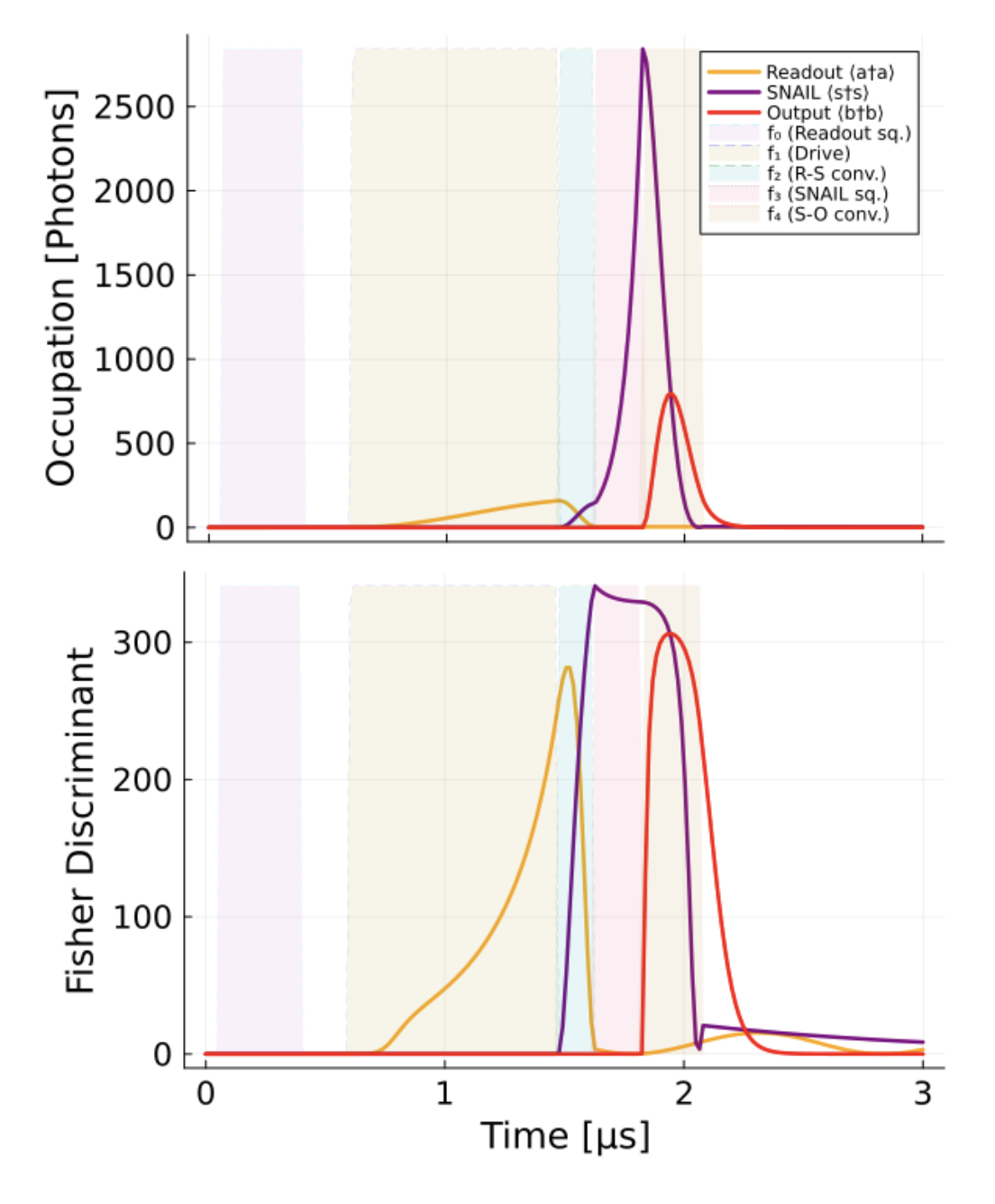}
    \caption{\textbf{(top)} The photon occupancies in the readout (yellow), SNAIL (purple) and output (red) throughout the readout sequence. \textbf{(bottom)} The SNR (or Fisher Discriminant) between the qubit state. The colored overlays indicate the pulses applied.}
    \label{fig:pop_and_snr}
\end{figure}

Our optimal sequence consists of five gates, illustrated in Fig. \ref{fig:seq_scheme}. 
\begin{enumerate}
    \item \textbf{Readout resonator squeezing} - Our first step is to prepare a squeezed vacuum state in the readout resonator. This allows us to decrease the noise, and thus improve fidelity without considerably increasing the number of photons in the readout resonator.
    \item \textbf{Readout resonator population} - The readout resonator is driven by a coherent microwave pulse and allowed to accumulate a phase due to the dispersive shift. 
    \item \textbf{Readout-SNAIL conversion} - The readout information is converted from the readout resonator into the SNAIL using a conversion pulse. The durations of the pulses up to this point are chosen such that the means are separated along the one axis, with their noise squeezed along the other axis. 
    \item \textbf{SNAIL Phase-sensitive amplification} - The process amplifies the signal along the separation axis. Since this amplification is phase-sensitive, there is no lower bound on the amount of noise added, and the amplification can be essentially noiseless.
    \item \textbf{SNAIL-Output conversion} - A final conversion process converts the SNAIL into the output, where the signal is coupled out to the rest of the readout chain via the fast output resonator.
\end{enumerate}

For the calculations and simulations in the manuscript, we assumed the effective couplings show in Table~\ref{tab:effective_rates}.

\begin{table}
    \centering
    \begin{tabular}{cc}\toprule
         Interaction& Coupling Rate [MHz]\\\midrule
         Readout Squeezing& 6.0\\
         Readout-SNAIL Conversion& 10.0\\
         SNAIL Phase-sensitive Amplification& 4.0\\
         SNAIL-Output Conversion& 10.0\\ \bottomrule
    \end{tabular}
    \caption{The effective quadratic coupling rates used in manuscript. Using the expected experimental parameters, these rates would correspond to driving rates of around 1-2GHz, and we expect that much higher rates could be used in practice.}
    \label{tab:effective_rates}
\end{table}
 
The evolution of the means and covariance due to each of these steps is given by Eq. \ref{eq:eoms}. In those terms, the different gates act as linear transformations in IQ space. In the rest of this section, we will follow in detail the different steps of the readout, their description, and effects on fidelity and information flow.

\subsection{Readout squeezing preparation}
To improve state discrimination without significantly increasing resonator occupation, we first apply a single-mode squeezing operation to the readout resonator. This would allow us to reduce the noise along a chosen quadrature in phase space, thereby enhancing distinguishability for a fixed signal strength.

We implement single-mode squeezing by driving the SNAIL with a pump at frequency \(\omega_p = 2\omega_a\), generating the effective Hamiltonian
\begin{equation}
    \hat{H}_{1} = {i r_a} \left( e^{-i\phi_a}\hat{a}^{\dagger 2} - e^{i\phi_a} \hat{a}^2  \right) 
\end{equation}
where $r_1$ is the effective strength of the squeezing interaction, determined by the strength of the pump, the hybridization strengths and SNAIL nonlinearity, as shown in Eqs. \ref{eq:effective_couplings}.  

In IQ space, in the basis of the quadrature amplitudes $\vec{X} = (\hat{q}_a, \hat{p}_a)^\intercal$, this interaction is generated by,
\begin{equation}
    \mathbf{\Omega H_{1}} = 
        \frac{r_a}{2} \begin{pmatrix}
            \sin\phi_a & -\cos\phi_a \pm \frac{\chi}{r_a} \\
            -\cos\phi_a \mp \frac{\chi}{r_a} & -\sin\phi_a
        \end{pmatrix}
\end{equation}
This transformation compresses the vacuum noise along one quadrature and amplifies it along the orthogonal one.  This form allows us to easily solve for the means and covariance matrix, and obtain the analytical expressions presented here.

\begin{figure}[h!]
    \centering
    \includegraphics[width=0.7\linewidth]{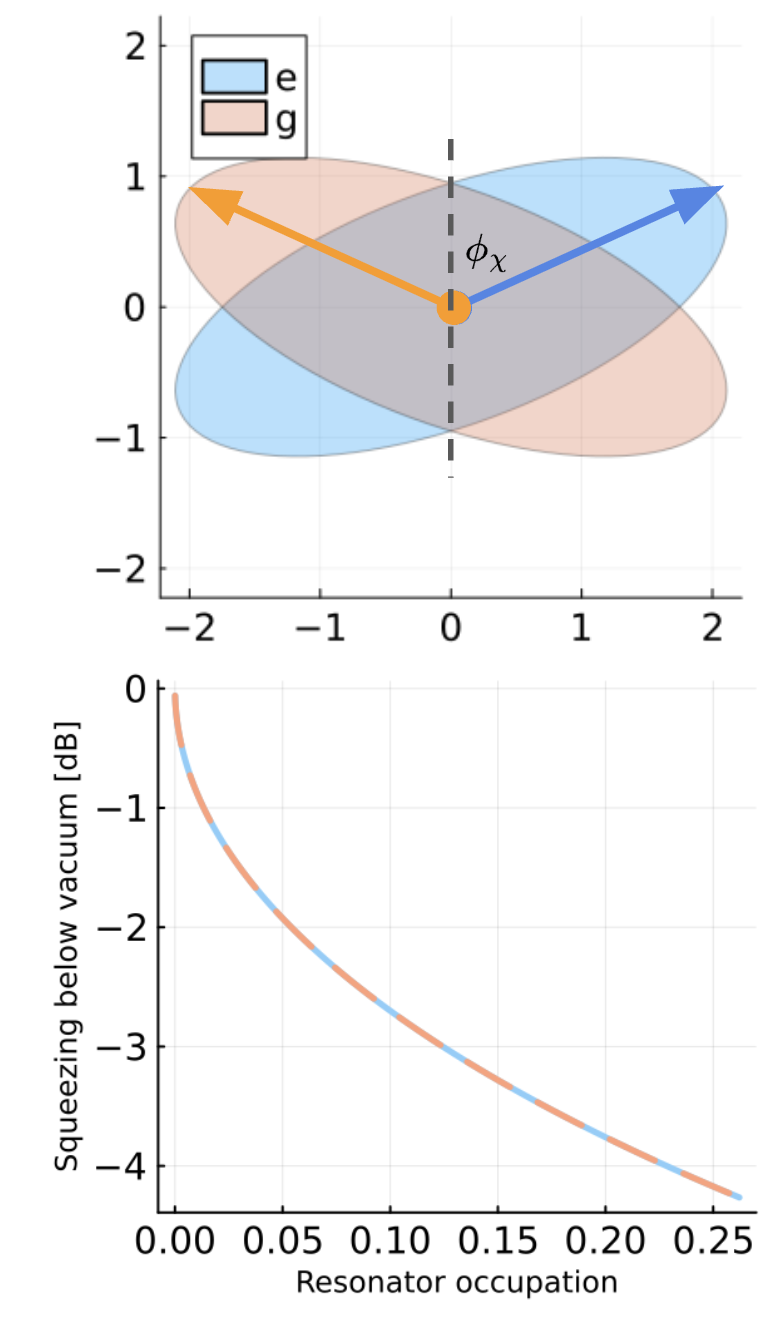}
    \caption{\textbf{top:} The IQ representation of the states for states detuned by $\pm\chi$. A delay of $\Delta T_1$ aligns the axes. \textbf{bottom:} Achievable squeezing (in dB) as a function of resonator photon number, showing constraints imposed by QND fidelity requirements.}
    \label{fig:sq_alignment_photons}
\end{figure}

Since the initial state is the vacuum state, the squeezing does not generate a change in the means, but rather just squeezes the noise of the system.
The optimal squeezing direction is orthogonal to the axis separating the final readout states in phase space. However, the squeezing interaction is inherently detuned due to the dispersive shift: depending on the qubit state, the resonator is shifted by $\pm\chi$, while the SNAIL is always driven at the bare resonance frequency.  This leads to an effective detuning, and consequently, the squeezing axes for the two logical states rotate in opposite directions.
\begin{equation}
    \tan \phi_\chi = \frac{1}{\chi \sinh r_\chi t} \left(r_\chi \cosh r_\chi t +  \sqrt{r^2 \cosh^2 r_\chi t - \chi^2}  \right)
\end{equation}
where we chose $\phi_a=0$, to simplify the expression. Note that a choosing a different phase simply acts to rotate both state in IQ-space, without affecting their relative phase. Additionally, we define $r_\chi = \sqrt{r^2 - \chi^2}$. Note that as long at the coupling rate $r$ is bigger than the dispersive shift $\chi$, we get fixed phase-sensitive amplification, albeit at a reduced rate.

This mismatch is generally small, but could lower our fidelity unless corrected. To resolve this, we can introduce a calibrated delay between the squeezing pulse and the displacement. This idle period allows the squeezed noise ellipses to rotate under their respective Hamiltonians until their axes align.
\begin{equation}
   T_{delay} = \frac{1}{2\chi}\left( \pi - 2\phi_\chi\right)
\end{equation}
which brings both squeezing axes into alignment perpendicular to the $Q$-axis.

Fig.~\ref{fig:sq_alignment_photons} (right) shows the evolution of the squeezing angles for the two logical states. After the delay, both ellipses rotate into alignment, ensuring that the squeezing noise suppression occurs along the correct axis for both readout outcomes.

An additional constraint on this step arises from the need to preserve the QND character of the readout. The squeezing operation populates the resonator with photons according to,
\begin{equation}
    n_a(t) = \frac{1}{2}\left( \left(\frac{r_a}{r_\chi}\right)^2  \cosh 2r_\chi t  - \left(\frac{\chi}{r_\chi}\right)^2  - 1\right)
\end{equation}
To avoid spurious qubit transitions or measurement-induced dephasing, we constrain the photon number to $n_a \le n_a^{\text{crit}} \sim 100$. Fig.\ref{fig:sq_alignment_photons} (left) shows the trade-off between squeezing strength (expressed in dB) and the photon budget. In our implementation, we find that moderate squeezing levels (e.g., 4-6 dB) are achievable while remaining within acceptable limits on resonator population.

This stage thus prepares the system for high-fidelity readout by engineering the quantum noise properties of the resonator, while preserving coherence and avoiding measurement backaction.

\subsection{Readout resonator population}
Following the squeezing step, we displace the readout resonator to imprint information about the qubit state  onto the field. 
\begin{equation}
    \hat{H}_{2} = \eta_d \left( e^{-i\theta_d} \hat{a} + e^{i\theta_d}\hat{a}^\dagger \right) 
\end{equation}
This is done by applying a coherent drive to the readout resonator at $\omega_d = \omega_a$, and allowing the system to evolve under the dispersive Hamiltonian $\hat{H}_\chi = \pm\chi \hat{a}^\dagger \hat{a}$. As the resonator is detuned by $\pm\chi$ depending on the qubit state, the resulting coherent states $|\alpha_{g (e)}(t)\rangle$ and rotate in opposite directions in phase space.

\begin{figure}[h!]
\centering
\includegraphics[width=0.85\linewidth]{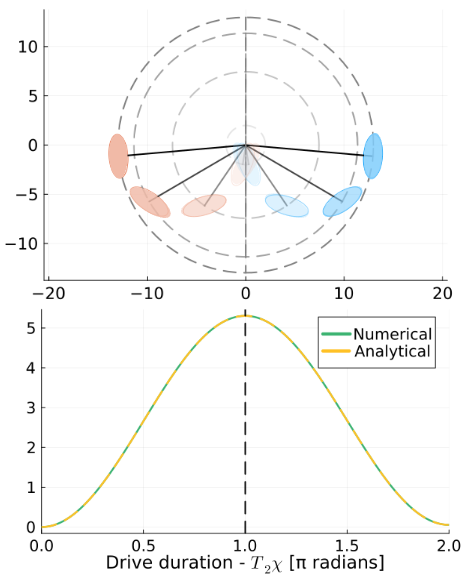}
\caption{\textbf{(top)} Coherent state trajectories $\alpha_{g(e)}(t)$ in IQ space, diverging due to dispersive rotation. \textbf{(bottom)} Separation $|\Delta\vec{\mu}_2(t)|$ as a function of time, peaking at $t = \pi/\chi$.}
\label{fig:readout_separation}
\end{figure}

The resulting separation $\Delta \vec{\mu}$ between the means of $|\alpha_{g (e)}(t)\rangle$ is given by,
\begin{equation}
    |\Delta\vec{\mu}_2| =    \frac{4 \eta_d }{\gamma_a^2 + 4 \chi ^2} \left(e^{-\frac{\gamma_a t}{2}} \left( \gamma_a \sin \chi t + 2 \chi  \cos \chi t \right) -2 \chi \right).
\end{equation}
This displacement traces a rotating trajectory in IQ space, as illustrated in Fig.~\ref{fig:readout_separation}.  This separation is maximized for time $\Delta T_2 =  \pi/\chi$, where the final separation lies along a single axis,
\begin{equation}
    |\Delta\vec{\mu}_2^*| = \frac{8 \chi \eta_d }{\gamma_a^2+4 \chi ^2} \left(e^{-\frac{\pi  \gamma_a}{2 \chi }} + 1 \right)
\end{equation}
The separation can grow unbounded with the drive rate, or by increasing the drive duration, but is limited by the critical readout occupation of the resonator.

Meanwhile, the covariance matrix continues to evolve under the dispersive shift. Thus, the noise ellipse remains squeezed but rotates with the coherent state. The time delay between the drive and the squeezing add a phase difference between the two that allows us to align them as required for maximal fidelity.

Fig.~\ref{fig:readout_separation} illustrates this behavior. The right panel shows the spiraling trajectories of the two coherent states, and the left panel shows the magnitude of their separation as a function of time. We choose the optimal duration $T_2$ to maximize distinguishability along the $q$-axis.

This stage completes the encoding of the qubit state into a measurable field displacement, while maintaining the noise suppression prepared by the prior squeezing step.

\subsection{Conversion into the embedded amplifier (SNAIL)}
At this stage, the readout resonator holds the readout information encoded in the displacement of its field. To prepare for amplification, we transfer this state into the SNAIL mode using a parametric conversion process. Unlike the readout resonator, the SNAIL can tolerate larger photon numbers, making it a suitable intermediary for amplification. Furthermore, since the interactions are not hybridized, the amplification can be performed efficiently even with modest pump strengths.

\begin{figure}[!h]
    \centering
    \includegraphics[width=0.6\linewidth]{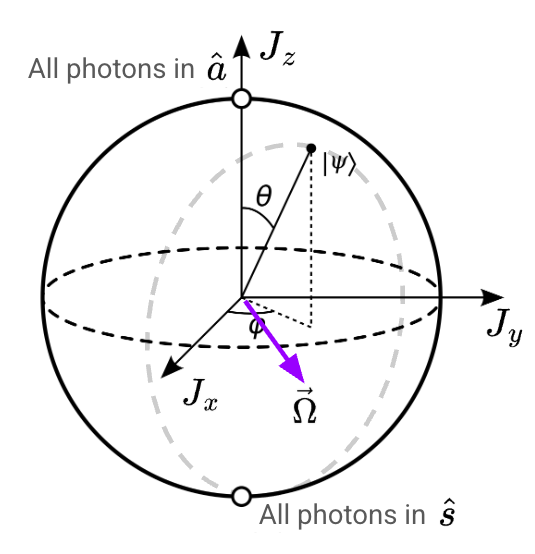}
    \caption{\textbf{Bloch sphere representation of resonator mode conversion.}  
The state of two coupled resonators with an arbitrary number of excitations map to a Bloch vector, with the poles corresponding to all photons in one resonator or the other. A resonant \(\pi\)-pulse (full iSWAP) drives equatorial rotation for full conversion. Detuning tilts the rotation axis, altering the trajectory and imprinting a relative phase on the final state.}
    \label{fig:bloch_sphere}
\end{figure}

We implement the conversion using a pump tone at the difference frequency between the readout resonator and the SNAIL $\omega_p = \omega_a - \omega_s$. This  hybridizes the readout resonator and the SNAIL, generating the interaction Hamiltonian,
\begin{equation}
    \hat{H}_{3} = \frac{1}{2}g_{as} \left( e^{i\theta_{as}} \hat{a}^\dagger \hat{s} + e^{-i \theta_{as}} \hat{a} \hat{s}^\dagger \right)
\end{equation}
where $\hat{a}$ and $\hat{s}$ are the annihilation operators for the readout and SNAIL modes, respectively, and $g_{as}$ is the effective conversion rate controlled by the pump.

The conversion Hamiltonian has an $SU(2)$ symmetry that describes dynamics identical to the precession of a spin in a magnetic field. It allows us to describe the dynamics in terms of generalized angular-momentum operators \cite{schwinger_angular_1952, yurke_su2_1986}, where the Hamiltonian describes Rabi-like precession around a generalized Rabi-vector,
\begin{equation}
    \vec{\Omega} = (g_{as}\cos \theta_{as}, -g_{as} \sin \theta_{as}, \pm\chi)
\end{equation}
in this space, the azimuthal angle describes the proportional division between the oscillators, and the equatorial angle describes the relative phase between the resonators, as illustrated in Fig.~\ref{fig:bloch_sphere}.

\begin{figure}[!h]
    \centering
    \includegraphics[width=\linewidth]{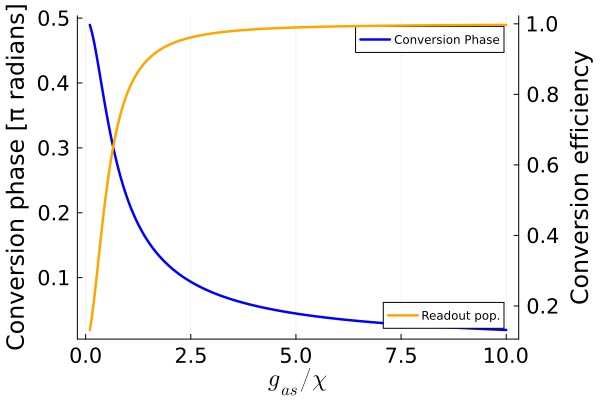}
    \caption{\textbf{(yellow)} Readout population after conversion. \textbf{(blue)} Additional conversion phase due to the detuning.}
    \label{fig:su2_conversion}
\end{figure}

In the absence of detuning, the state fully transfers from one mode to the other over a time $T_\pi = \pi / (2 g_\text{as})$. However, the dispersive shift $\pm \chi$ introduces an effective detuning, modifying the Rabi rate $|\Omega| = \sqrt{g_\text{as}^2 + \chi^2}$.

\begin{equation}
    p_a = \frac{|\chi|^2}{|g_{as}|^2 + |\chi|^2} \qquad p_s = \frac{|g_{as}|^2}{|g_{as}|^2 + |\chi|^2} 
\end{equation}
The maximum transfer amplitude is then reduced, and the conversion acts as a partial beam splitter unless $g_\text{as} \gg \chi$. This limits the efficiency of processes such as frequency conversion. In particular, because the frequency of the readout resonator is shifted by \(\pm \chi\), depending on the qubit state, the conversion process is inherently off-resonant by an amount \(\chi\), resulting in limited efficiency. This residual excitation, left behind due to incomplete transfer, acts as a continuous weak measurement on the qubit, leading to decoherence. As shown in Fig.\ref{fig:su2_conversion}, this efficiency can increase arbitrarily using stronger parametric couplings. To suppress these effects further, additional conversion gates, or readout displacement pulses, can be added. Although this does not change conversion efficiency, it could be used to empty the readout resonator and prevent measurement-induced decoherence. Additionally, more complex techniques such as composite pulses could be employed. These are not necessarily more sophisticated, but are details that the experimental implementation would have to address.

In addition to limiting transfer fidelity, the detuning introduces an additional phase accumulation during the conversion. 
\begin{equation}
    \tan \theta_\chi = \pm \frac{\chi}{|\Omega|}
\end{equation}
This phase needs to be accounted for to ensure the final state remains aligned along the measurement axis.

As illustrated in Fig.~\ref{fig:su2_conversion}, this corresponds to a partial spin rotation on the Bloch sphere defined by the two-mode Hilbert space. The detuning-induced phase $\theta$ must be compensated in the subsequent amplification step to maintain alignment of the state along the optimal readout axis.

To ensure high-fidelity transfer, we operate in the regime $g_{as} \gg \chi$, which suppresses both transfer loss and excess phase accumulation. This conversion step thus moves the readout state into the amplification-ready SNAIL mode, preserving its squeezed, phase-sensitive character.

Importantly, this is the last step that interacts directly with the dispersively shifted readout mode. It therefore defines the maximum achievable readout fidelity. Subsequent stages can amplify and extract the signal, but cannot improve the fundamental distinguishability between the readout states.

\subsection{Phase-sensitive amplification} 
After transferring the readout signal into the SNAIL mode, we perform phase-sensitive amplification along the axis of maximal signal separation. This amplification enhances the signal-to-noise ratio (SNR) without introducing additional noise along the measurement quadrature.

We implement this step using a parametric pump at twice the SNAIL frequency, generating the squeezing Hamiltonian:
\begin{equation}
    \hat{H}_4 = ir_s \left( e^{-i\phi_s} \hat{s}^{\dagger 2} - e^{i\phi_s} \hat{s}^2 \right),
\end{equation}
where $r_4$ is the gain rate and $\phi_4$ sets the amplification axis. By aligning $\phi_s$ with the signal quadrature (typically the $Q$-axis), we achieve amplification with no added noise along that axis.
The phase of the amplification must be chosen to lay along this axis,
\begin{equation}
    \phi_{opt} = \theta_d + \pi/2
\end{equation}
Note that the IQ representation is identical to the readout squeezing, except that this process is no longer detuned.

\begin{figure}[!h]
    \centering
    \includegraphics[width=0.85\linewidth]{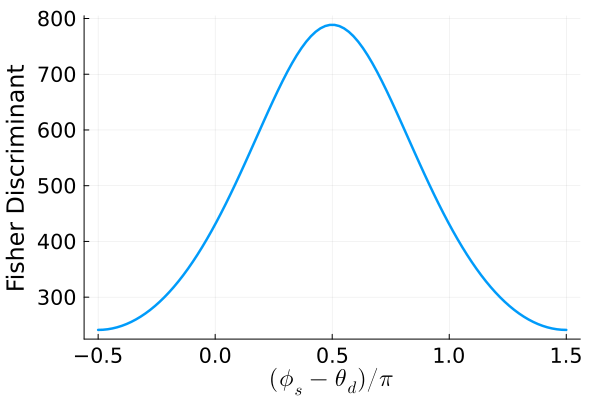}
    \caption{SNR (of Fisher Discriminant) as a function of the amplification phase.}
    \label{fig:amp_fd}
\end{figure}

This is crucial for downstream processing. According to Friis' formula \cite{friis_noise_1944}, the effective noise added by subsequent elements in the amplification chain is suppressed by the gain of the first amplifier. By performing strong, quantum-limited amplification at this stage, we ensure that the rest of the readout chain does not degrade the distinguishability of the readout states.

Importantly, because the readout states have been engineered to lie along a single quadrature, phase-sensitive amplification is not only sufficient but optimal. The process preserves the covariance structure and enhances separation without distorting the information content.

This step completes the signal preparation for final extraction, ensuring that all relevant quantum information has been captured with minimal added noise.

\subsection{Conversion to the output} 
After phase-sensitive amplification, the readout information is encoded in a large, coherent displacement in the SNAIL mode. The final step in the readout protocol is to transfer this amplified signal into a fast, low-Q output resonator coupled to the external environment, where it can be detected with minimal distortion.

\begin{figure}[!h]
    \centering
    \includegraphics[width=0.85\linewidth]{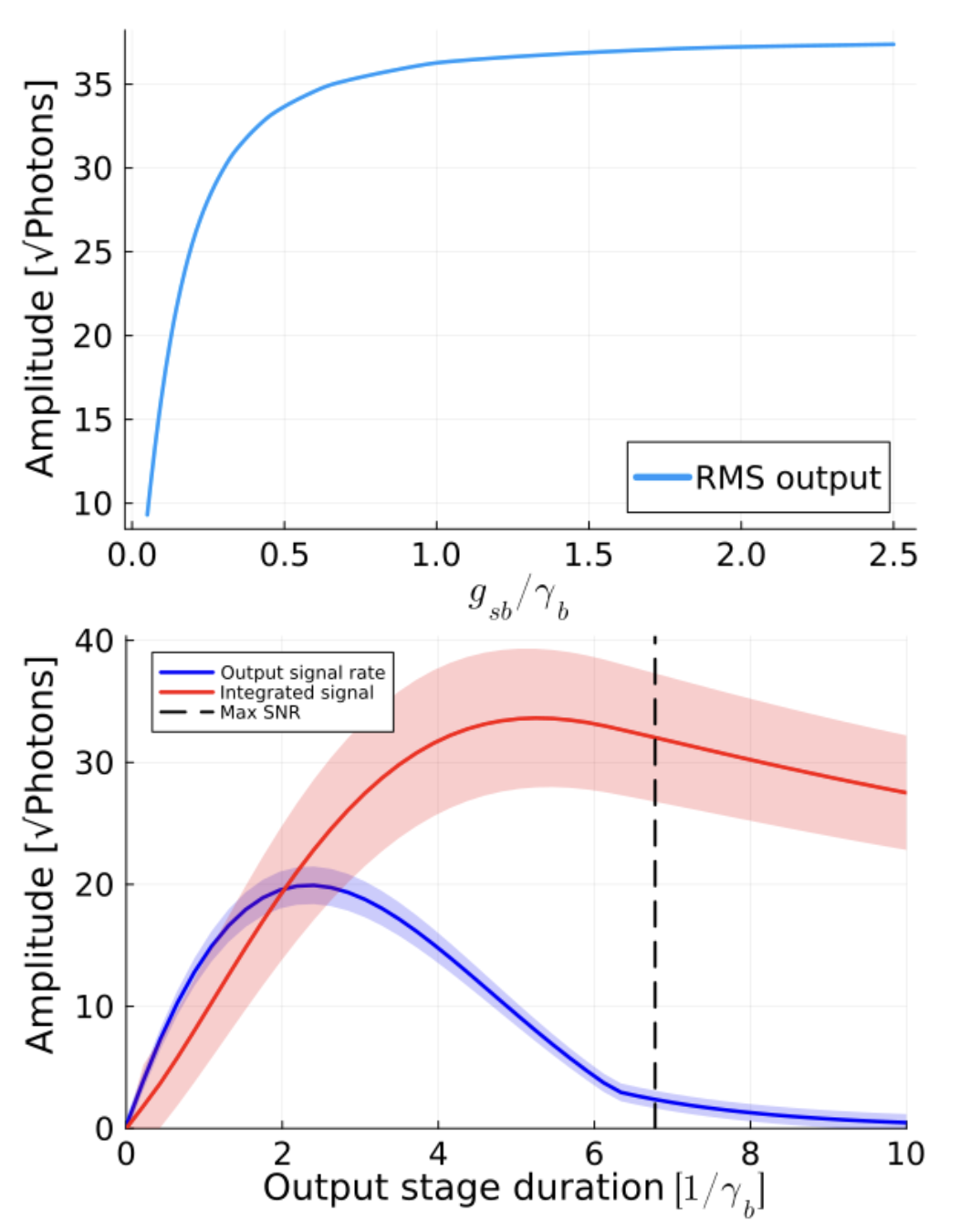}
    \caption{\textbf{(top)} The output photon rate as a function of $g_{sb}/\gamma_b$; the conversion rate efficiency increases as it approaches $\gamma_b$. Due to the output decay, the state loses coherence and the conversion becomes less effective. To ensure optimal conversion, the conversion rate $g_{sb}$ must be comparable to the output decay rate $\gamma_b$. \textbf{(bottom)} The intracavity output signal (in blue), and the integrated signal at the output (in red).  The SNR peaks once the photons leave the output resonator.}
    \label{fig:final_conversion}
\end{figure}

We implement this transfer using another parametric conversion pulse, tuned to the frequency difference $\omega_p = \omega_b - \omega_s$ between the SNAIL and the output resonator:
\begin{equation}
    \hat{H}_{5} = \frac{1}{2}g_{sb}(e^{i\theta_{sb}} \hat{s}^\dagger \hat{b} + e^{-i\theta_{sb}} \hat{b}^\dagger \hat{s})
\end{equation}
where $\hat{s}$ and $\hat{b}$ are annihilation operators for the SNAIL and output modes respectively, and $g_{sb}$ is the tunable conversion rate. This is very similar to the readout-SNAIL conversion, but with two important differences: \textbf{(1)} it is no longer detuned, since we know exactly the frequency difference between the SNAIL and the output resonators, and \textbf{(2)} The output is leaky, making the interaction rapidly lose coherence. Note that although at this point vacuum fluctuations leak into the signal, they are of no consequence because of the SNAIL amplification. 

Because the output resonator is engineered to be strongly coupled to a transmission line (with decay rate $\kappa_b$), the transferred signal rapidly leaks into the external measurement chain. 

As no further transformations are performed on the signal, the distinguishability of the output states is entirely determined by the accumulated amplification and separation up to this point. We therefore interpret this stage as a “readout release” gate: it does not modify the information content, but rather transfers it to the measurement apparatus. 

The noise of the signal is obtained similarly by applying input-output relations to the signal variance. 
The intra-cavity signal, and the output integrated signal, as well as their noise, are plotted in Fig. \ref{fig:final_conversion}.

\section{Output Fidelity Analysis}
\label{sec:output_fidelity}
In the previous sections, we optimized the Embedded Amplifier (EA) readout protocol by jointly tuning the pulse sequence and parametric drive strength to enhance signal separation while minimizing residual resonator occupation.

The EA scheme offers several fundamental advantages over Conventional Dispersive Readout (CDR):
\begin{enumerate}
    \item \textbf{Maximized Information Efficiency:} By \emph{temporally decoupling} information accumulation from signal extraction, the EA protocol maximizes the Signal-to-Noise Ratio (SNR) extracted per photon, utilizing the full cavity lifetime to acquire phase information \emph{before} release. 
    
    \item \textbf{Enhanced Quantum Efficiency:} The integration of an on-chip phase-sensitive amplifier enables \emph{noiseless amplification} of the signal at the source. This architecture boosts the signal and permits the elimination of lossy external components, thereby significantly improving the total quantum efficiency of the measurement chain. 
    
    \item \textbf{Back-Action Evasion via Squeezing:} The scheme naturally accommodates squeezed states, allowing measurement noise to be reduced below the Standard Quantum Limit, allowing for high-fidelity readout with reduced intracavity photon numbers.
\end{enumerate}

In this section, we assess the overall performance of the EA scheme and benchmark it against CDR, highlighting the practical advantages enabled by this "Catch-Process-Release" architecture.

At a fundamental level, both EA and CDR rely on the same physical encoding: qubit-state information is mapped onto the quadrature amplitudes of a resonator via the dispersive interaction. However, the \emph{timing} of how this information is encoded and released differs fundamentally.

We can illustrate this by comparing the evolution of the field quadratures. For CDR, the system is driven continuously, and the field evolves towards a steady state while simultaneously leaking into the transmission line:
\begin{align}
\langle \hat{X} \rangle^{(\mathrm{CDR})}
&= \sqrt{\overline{n}}
\Big[
\left(1 - e^{-\kappa t/2}\cos(\chi t)\right)\cos\phi
\nonumber\\
&\hspace{2em}
- e^{-\kappa t/2}\sin(\chi t)\,\sin\phi
\Big],
\qquad t < T
\\[0.5em]
\langle \hat{Y} \rangle^{(\mathrm{CDR})}
&= \sqrt{\overline{n}}
\Big[
\left(1 - e^{-\kappa t/2}\cos(\chi t)\right)\sin\phi
\nonumber\\
&\hspace{2em}
+ e^{-\kappa t/2}\sin(\chi t)\,\cos\phi
\Big],
\qquad t < T .
\end{align}
where $T$ is the pulse duration, with free decay for $t > T$. We defined the standard "mean occupation" $\overline{n} = \frac{\kappa |\beta_0|^2}{(\kappa/2)^2 + \chi^2}$ and steady-state phase shift $\phi = \arctan\left(\frac{2\chi}{\kappa}\right)$.

Crucially, in CDR, the information extraction is continuous. Photons emitted at early times $t \ll 1/\kappa$ carry almost no information. Yet, these "premature" photons contribute to the shot noise of the measurement just as much as later photons, diluting the overall SNR. 

In contrast, the EA scheme adopts a "Catch-Process-Release" protocol that temporally separates accumulation from extraction. The output field during the release window is given by:

\begin{align}
\langle \hat{X} \rangle^{(\mathrm{EA})}
&= G\,s(t)\,\sqrt{\overline{n}}
\Big[
\cos\phi
- e^{-\gamma_a t_a/2}\cos(\chi t_a)\cos\phi
\nonumber\\
&\hspace{2em}
- e^{-\gamma_a t_a/2}\sin(\chi t_a)\sin\phi
\Big]
\\[0.5em]
\langle \hat{Y} \rangle^{(\mathrm{EA})}
&= G^{-1}s(t)\,\sqrt{\overline{n}}
\Big[
\sin\phi
- e^{-\gamma_a t_a/2}\cos(\chi t_a)\sin\phi
\nonumber\\
&\hspace{2em}
+ e^{-\gamma_a t_a /2} \sin(\chi t_a)\cos\phi
\Big].
\end{align}
Here $s(t) = a(\Delta t)e^{-\gamma_b \Delta t/2}$ is the temporal envelope of the released burst (with $a(t)$ turning on at $t_{proc}$ with $\Delta t = t - t_{proc}$), and $G$ is the amplification gain.

\begin{figure}
    \centering
    \includegraphics[width=0.85\linewidth]{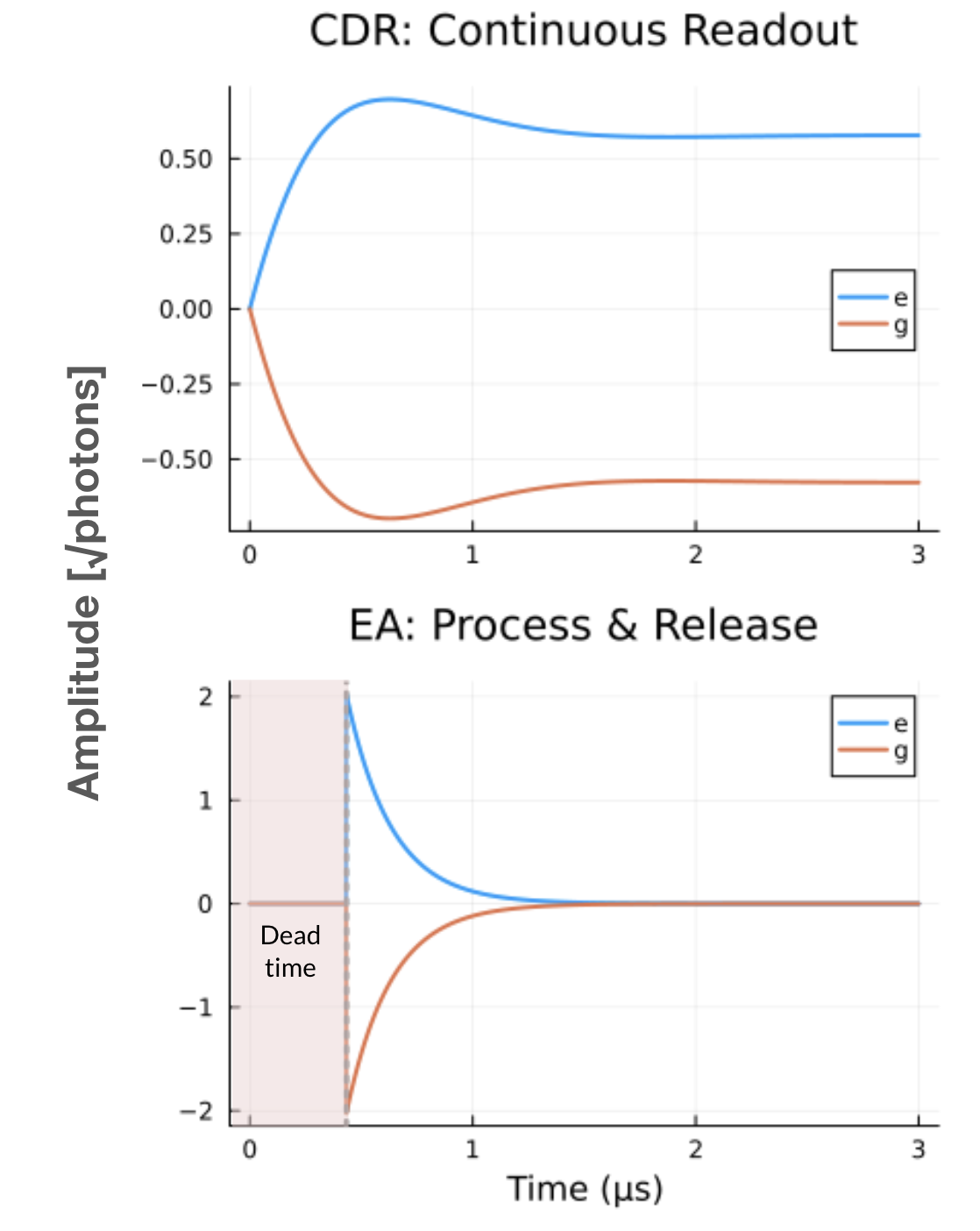}
    \caption{\textbf{Readout field dynamics.} \textbf{(a)} CDR: The resonator is driven continuously towards a steady state. The transient oscillation arise from the rotation of the field quadratures during signal accumulation. \textbf{(b)} EA: The "Catch-Process-Release" protocol accumulates signal in a high-Q cavity before releasing it in a compressed burst (dashed line). This temporal compression yields a peak amplitude significantly higher than the CDR steady state, maximizing instantaneous SNR even without active amplification. In practice, the rise time would depend on the conversion pump strength.}
    \label{fig:ea_cdr_traj}
\end{figure}

The key difference lies in the arguments of the trigonometric terms. In EA, the accumulated phase $\chi t_a$ is \textbf{fixed} and maximized before the release begins. Every photon in the output pulse carries the full phase-information accumulated over the entire interaction time $t_a$. This separation ensures that no photon is wasted; the "information density" per photon is maximized before the signal is ever exposed to the noise of the transmission line. Furthermore, the on-chip amplification $G$ allows us to amplify the information before it reaches the transmission line, making sure it is unaffected by the readout chain noise.

\subsection{Measurement Statistics \& SNR}
The intracavity fields are not physically accessible; instead, the experiment records a noisy current trace at the output of the amplification chain. Using standard input-output theory, the homodyne current operator $\hat{I}(t)$ is given by~\cite{wiseman_quantum_2009, clerk_introduction_2010}:
\begin{equation}
    \hat{I}(t) = \sqrt{2 \eta \gamma_{meas}} \hat{X}(t) + \hat{\xi}(t)
\end{equation}
where $\eta$ is the total collection efficiency, $\gamma$ is the measurement rate ($\kappa$ for CDR and $\gamma_b$ for EA). The operator $\hat{\xi}$ represents the effective noise, and it includes both the vacuum fluctuations entering the detection chain and the added noise of the amplifiers:
\begin{equation}
    \langle \hat{\xi}(t) \hat{\xi}(t')\rangle = \delta(t-t') \left (\frac{1}{2} + n_{add} \right)
\end{equation}

\textbf{Efficiency Advantage:} A critical distinction between the protocols lies in the attainable efficiency $\eta$. In CDR, the output signal must be routed through a series of circulators and isolators to protect the qubit from back-action noise. These components introduce insertion losses that directly degrade the quantum efficiency. Typical insertion losses are around $0.3-0.5\mathrm{dB}$, leading to efficiencies of $\eta_{CDR} \approx 0.7 - 0.8$ \cite{noauthor_lnf-xxxxc4_8b_2025}.

In contrast, the EA architecture integrates a \emph{phase-sensitive} parametric amplifier directly on-chip. This provides noiseless amplification ($n_{add} \rightarrow 0)$ \emph{before} the signal encounters any off-chip losses. Furthermore, the directional nature of the EA protocol allows for the elimination of bulk circulators, significantly reducing routing losses. Consequently, the effective efficiency can approach unity, $\eta_{EA} \approx 1$, limited by the inefficiency of the conversion process which can be mitigated by increasing the conversion pump strength, or by using more involved pulse schemes.

\textbf{Integrated Statistics:} Since the readout dynamics are linear and the noise is Gaussian, the complete information for state discrimination is contained within a single scalar statistic, the integrated current $\hat{Z}$.
\begin{align}	
	&\hat{Z} \equiv \int_0^T g(t) \hat{I}(t) dt 
\end{align}
where $g(t)$ is a temporal filter function, typically chosen to be normalized $\int_0^T g^2(t)dt = 1$. The integrated traces are often used as discrimination and visualization statistic. Due to their experimental utility and simplicity, we will use them as our decision statistic and the basis for our fidelity calculations and comparison. However, we note that although they are often used even when the underlying processes are not Gaussian, in that case the integrated traces are sub-optimal, and maximal fidelity requires using the entire trace. 

As in the previous sections, the performance of the measurement is quantified by the squared SNR:
\begin{equation}
    	\mathcal{D}^2 = \frac{(\mathbb{E}[Z \,|\, e] - \mathbb{E}[Z \,|\, g])^2}{\mathrm{Var}[Z]}
\end{equation}
Generally, the noise (variance) of the integrated current will have two contributions:
\begin{align}
\text{Var}[\hat{Z}(T)]
&= \underbrace{\left(\frac{1}{2} + n_{add} \right)\int_0^T g^2(t)\,dt}_{\text{shot-noise}}
\nonumber\\
&\quad
+ \underbrace{
2\eta\gamma
\iint_0^T dt_1\,dt_2\,
g(t_1)g(t_2)\,
\Gamma_{XX}(t_1,t_2)
}_{\text{intra-cavity fluctuations}} .
\end{align}
where $\Gamma_{XX}(t_1, t_2)$ describes the correlations of the intracavity field quadrature. The first term is the detector shot-noise, while the second term is the noise due to the intracavity fluctuations. The on-chip amplification allows us to drown the shot-noise, and be only limited by the intracavity fluctuations.

To maximize the SNR in the presence of white noise, we employ the optimal \textbf{matched filter}, which weighs the time-dependent signal contrast $w_{opt}(t) \propto \langle\hat{I}_e \rangle - \langle\hat{I}_g \rangle := \Delta I$. In more general Gaussian cases—e.g., with correlated or time-dependent noise—the optimal filter can have a nontrivial temporal structure that depends explicitly on the covariance matrices of the measurement processes~\cite{helstrom_statistical_1968, khan_practical_2023}.

\begin{figure}
    \centering
    \includegraphics[width=0.85\linewidth]{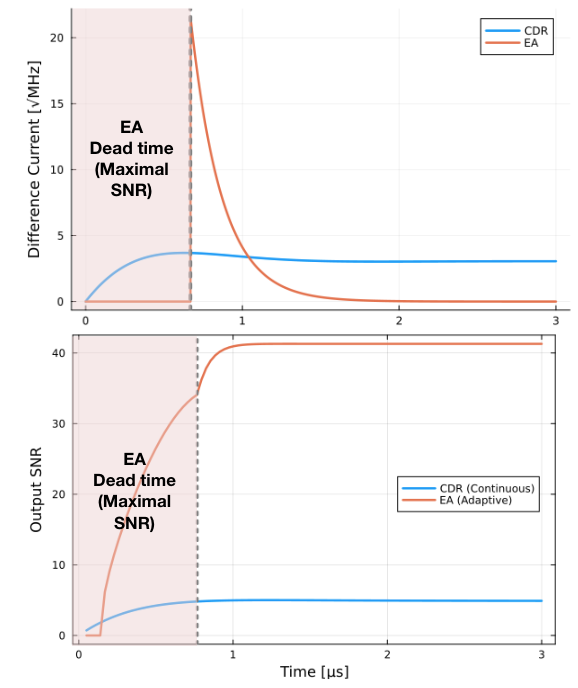}
    \caption{\textbf{Instantaneous signal contrast and SNR accumulation.} 
    \textbf{(a)} Comparison of the measurement signal rate $|\Delta I(t)|$ (difference current) for the CDR and EA protocols. While CDR (blue line) extracts information as a continuous, low-amplitude stream, the EA protocol (orange line) accumulates information in the cavity before releasing it in a compressed, high-intensity burst at $t \approx 1\,\mu\text{s}$. This temporal compression yields a peak signal amplitude significantly higher than the CDR steady state. 
    \textbf{(b)} Accumulated Signal-to-Noise Ratio (SNR) as a function of the total readout time $T_{total}$. The EA protocol (orange) incurs an initial ``dead time'' penalty (approx. $50 \text{ns}$) for pulse overheads, but the SNR rises rapidly once the release phase begins, quickly surpassing the linear accumulation of the CDR protocol (blue). The EA SNR saturates once the optimal interaction time ($t_a = \pi/2\chi$) is reached, indicating that the maximal information content of the photon budget has been successfully extracted. The calculated example uses a photon budget of $5.0$.}
    \label{fig:snr_vs_t}
\end{figure}

Using this optimal filter, the accumulated Power SNR is given by the integral of the instantaneous signal power normalized by the noise variance. To ensure a fair resource comparison, we impose a global time constraint $T_{total}$ on both protocols. For CDR, this includes the readout drive duration and the integration time. For the EA protocol, since the total duration has multiple contributions from different stage, the pulse sequence is adaptively resized: the interaction time $t_a$ is expanded to fill the available budget $T_{total} - t_{dead}$ until the optimal interaction time is reached, where $t_{dead}$ accounts for the fixed overheads of the conversion, and amplification pulses.

Figure~\ref{fig:snr_vs_t} illustrates the SNR accumulation as a function of the total time constraint. The results highlight the main advtange of the EA protocol: even at unity gain ($G=1$), the EA protocol outperforms CDR after overcoming the initial dead-time penalty. This advantage arises from the separation of the information accumulation and extraction. By ``catching'' the photons, letting them interact with the readout resonator, and releasing them in a short burst once the information has been optimally accumulated, the EA scheme maximizes the SNR per photon.

Crucially, the EA SNR saturates once the interaction time reaches the optimal limit $t_a = \pi/2\chi$. At this point, the maximum possible phase shift has been acquired; extending the time further yields no additional information, whereas CDR requires significantly longer integration times to asymptotically approach the same information content, although it is able to produce information as soon as the readout pulse starts.

\begin{figure}
    \centering
    \includegraphics[width=0.85\linewidth]{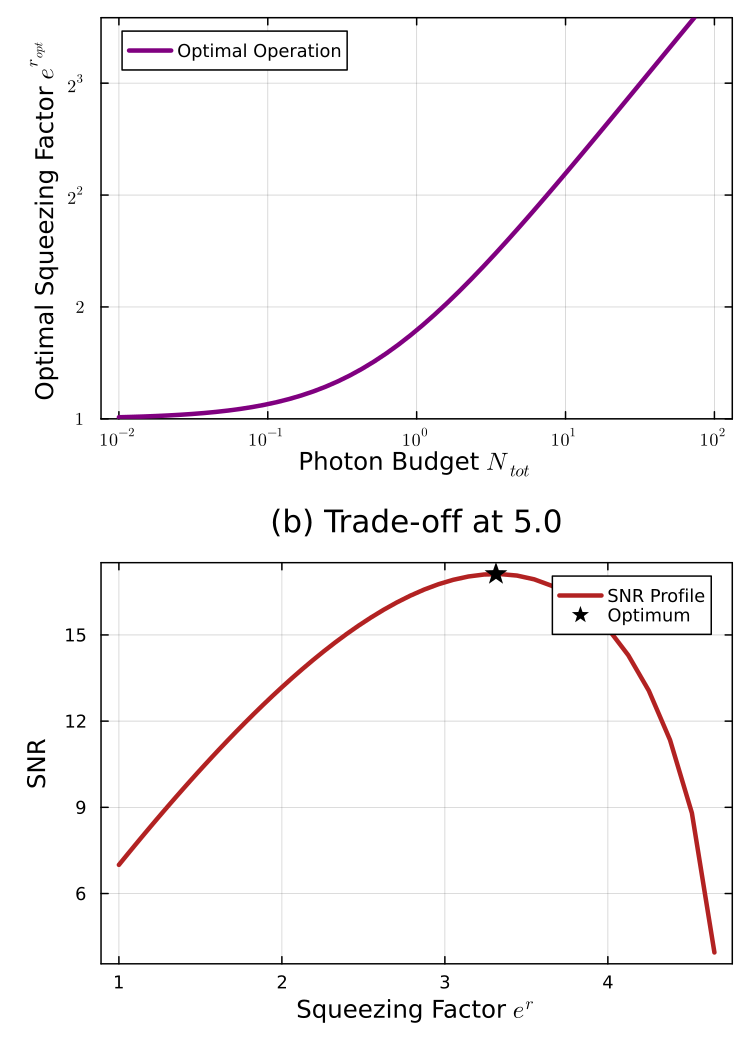}
    \caption{\textbf{Resource dependence of the Output SNR.} 
    \textbf{(a)} The optimal amount of squeezing for maximal SNR as a function of the available photon budget. As the squeezing increases, the number of photons available for displacing the distribution decreases, resulting in a well-defined optimum.
    \textbf{(b)} SNR versus squeezing factor in the high-gain limit ($G=100$). Squeezing suppresses vacuum fluctuations, boosting SNR beyond the standard quantum limit. The eventual roll-off and collapse indicate the depletion of the photon budget, where the energy cost of generating the squeezed vacuum consumes the capacity needed for the coherent signal field.}
    \label{fig:snr_vs_sq}
\end{figure}

Furthermore, the performance of the EA protocol is governed by two distinct resource thresholds: the amplification necessary to overcome added noise, and the squeezing constrained by the overall photon budget. As shown in Fig.~\ref{fig:snr_vs_sq}(a), increasing the parametric gain $G$ initially yields a sharp improvement in SNR by lifting the signal variance above the constant background of the following detection chain. However, once the added noise is effectively diluted ($G \gg 1$), the system enters a quantum-limited regime where further gain offers no advantage, as the vacuum fluctuations are amplified alongside the signal. In this saturated regime, applying squeezing becomes the primary lever for enhancement (Fig.~\ref{fig:snr_vs_sq}(b)). By suppressing the vacuum noise along the measurement quadrature, squeezing enables SNR values exceeding the standard quantum limit. Crucially, this benefit is not unbounded; the generation of highly squeezed states requires a significant intracavity photon population ($N_{sq} \propto \sinh^2 r$). As the squeezing strength increases, this ``vacuum overhead'' eventually cannibalizes the fixed total photon budget $N_{tot}$, leaving insufficient energy for the coherent signal displacement and causing the SNR to collapse. 

\subsection{Fidelity and Resource Benchmarking}

Ultimately, the figure of merit for qubit readout is the assignment fidelity $\mathcal{F} = 1 - (1-p_{mis})(1 - p_{rel})$, where $p_{mis} = \text{erfc} \bigg(\frac{\mathcal{D}}{2\sqrt{2}} \bigg)$ is the classification error and $p_{rel} = 1 - e^{T/T_1}$ is the qubit-relaxation error. This metric balances the classification error, which decreases with photon number and time, against the relaxation error $p_{rel}$, which increases with time as the qubit decays.

\begin{figure}[!h]
    \centering
    \includegraphics[width=0.75\linewidth]{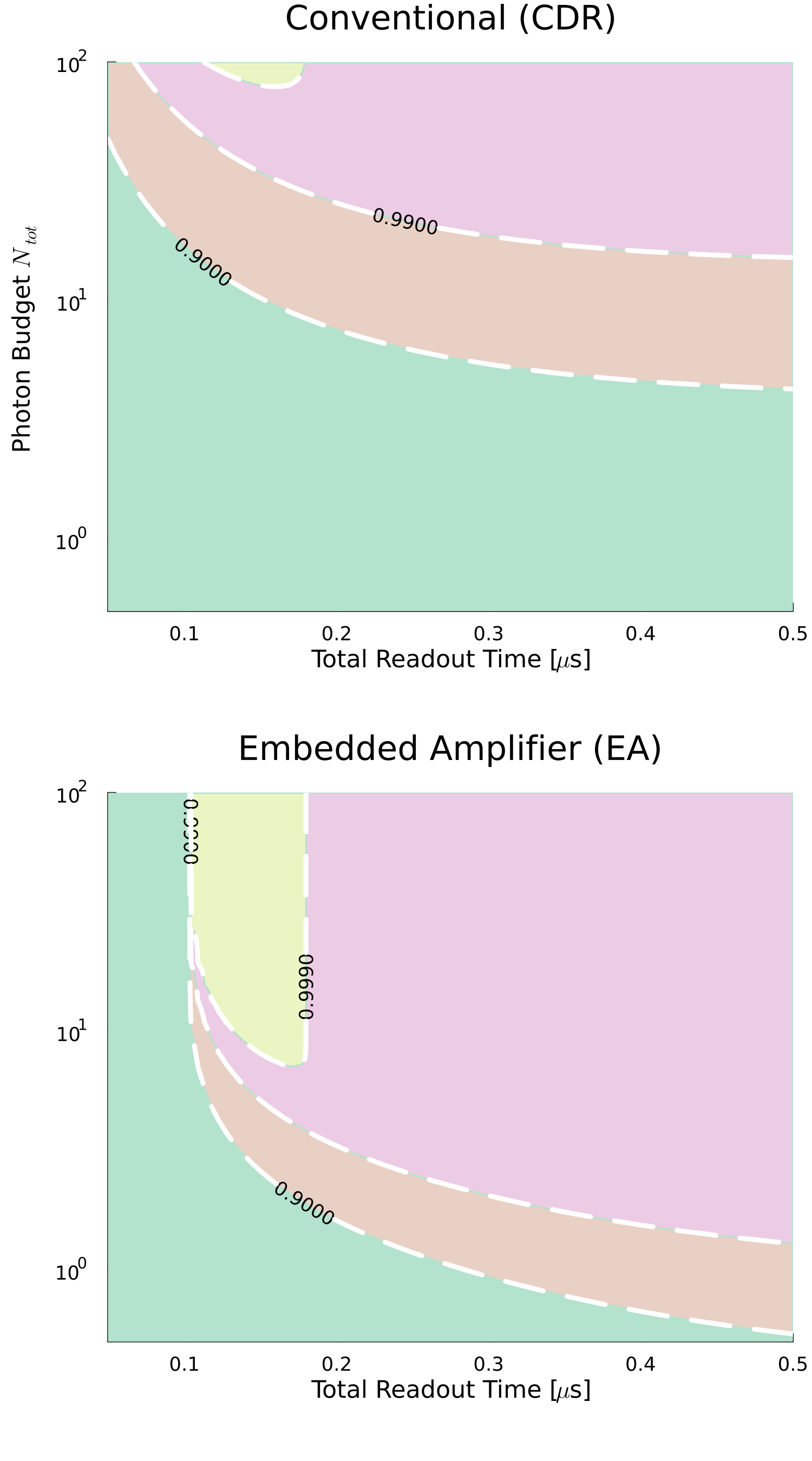}
    \caption{\textbf{Fidelity contours vs.\ resource constraints.} Readout fidelity (in ``nines'', $-\log_{10}(1-\mathcal{F})$) plotted against photon budget $N_{tot}$ and total time $T$. \textbf{(Top) CDR:} The high-fidelity region is strictly limited by the trade-off between signal accumulation rate and qubit decay. The gray dotted line is break-even line of the two schemes, indicating that CDR may still be preferable when very fast readout is required. \textbf{(Bottom) EA:} Despite the initial dead time ($t_{dead} \approx 0.3\,\mu\text{s}$), the protocol achieves 3-nines (99.9\%) and 4-nines benchmarks (purple/yellow regions) with significantly fewer photons and shorter total times than CDR, demonstrating the superior efficiency of the catch-process-release scheme. }
    \label{fig:fidelity_contours}
\end{figure}

As common in the field, we define the readout ``infidelity'' in terms of the number of nines (e.g., $10^{-3}$ corresponds to $99.9\%$). Figure~\ref{fig:fidelity_contours} compares the performance boundaries of both protocols over a wide parameter space of total photon budget $N_{tot}$ and total time constraint $T$.

\textbf{Conventional Dispersive Readout (CDR):} The CDR performance (Fig.~\ref{fig:fidelity_contours}a) exhibits the characteristic ``hill'' topology defined by the competition between SNR and decay. 
Achieving high fidelity ($>99.9\%$) requires a delicate balance: the time must be long enough to integrate the signal but short enough to avoid $T_1$ decay ($T \ll T_1$). As a result, the high-fidelity region is narrow and bounded.

\textbf{Embedded Amplifier (EA):} The EA performance (Fig.~\ref{fig:fidelity_contours}b) displays a fundamentally different scaling behavior:
\begin{itemize}
    \item \textbf{The Dead Zone:} The gray region at $T < 300\,\text{ns}$ represents the protocol overheads. During this window, no information can be extracted, setting a hard speed limit on the measurement.
    \item \textbf{Steep Ascent:} Once $T > t_{dead}$, the fidelity rises sharply, since each readout photon contains the maximal amount of information.
    \item \textbf{Expanded High-Fidelity Region:} The contours extend over a much broader range of photon budgets compared to CDR. Notably, the EA protocol achieves the $99.9\%$ benchmark with significantly fewer photons.
\end{itemize}

This analysis highlights a fundamental constraint of the EA architecture: the fidelity is asymptotically bounded by qubit decay during the dead time. Even in the limit of infinite SNR, the maximum achievable fidelity is capped at $\mathcal{F}_{max} \approx e^{-t_{dead}/T_1}$. While this ceiling remains well above $99.9\%$ for the parameters considered here ($T_1 = 50\,\mu\text{s}$), it underscores a critical design trade-off. Although CDR utilizes the photon budget far less efficiently, it incurs zero temporal overhead, allowing information accumulation to commence immediately upon the application of the readout drive.

\section{Robustness and Scalability}
\label{sec:non-idealities}
While the hybridized SNAIL platform offers a promising path toward compact, coherent quantum-state readout, its practical implementation raises a number of robustness and scalability considerations. In real-world settings, non-idealities such as parameter drifts, residual nonlinearities, and inter-resonator crosstalk can influence performance. Additionally, as the architecture is extended to support multiple qubits, constraints related to frequency crowding, control complexity, and calibration overhead become increasingly relevant. While a full characterization of these effects is beyond the scope of this work, and would delicately depend on experimental details, we outline the key challenges and tradeoffs that arise in realistic scenarios, and highlight considerations important for scaling this platform.

\qquad
 
\emph{Sensitivity to parameter variations} -- The readout fidelity depends on precisely tuned interactions - particularly in the sequential protocol, where timing and rate control must be accurate relative to the small dispersive shifts. Modest drifts in pump amplitude, detuning, or squeezing strength can lead to reduced state separation or incomplete transfer to the SNAIL mode. While current experimental platforms can meet these timing requirements, the demands may become more stringent in systems with larger dispersive couplings or stronger nonlinearities. 

\qquad

\emph{Parametric speed-limits and conversion efficiency} -- While our linearized model suggests that the coupling strength \( g_{\omega_i \omega_j} \) between modes can be tuned arbitrarily, in practice, this is fundamentally constrained by the nonlinear nature of the SNAIL. As the pump power increases, the SNAIL mode becomes increasingly populated, eventually entering regimes of strong nonlinearity, saturation, and even chaotic dynamics. These effects significantly alter the system's behavior and set a practical upper limit on the usable coupling strength, typically on the order of a few MHz. This constraint imposes a fundamental "speed limit" on parametric gate operations \cite{xia_exceeding_2025}.

\qquad

\emph{Residual Kerr non-linearities} -- SNAIL devices are typically biased at a Kerr null point, where the static Kerr nonlinearity is minimized to suppress unwanted frequency shifts. However, residual Kerr effects persist due to imperfect cancellation and additional dynamical contributions, such as counter-rotating terms beyond the rotating-wave approximation (RWA)~\cite{bello_systematic_2025, miano_frequency-tunable_2022}. This residual nonlinearity manifests primarily as a self-Kerr effect in the SNAIL mode, with weaker cross-Kerr couplings to other resonators. Self-Kerr can introduce significant photon-number-dependent frequency shifts in the SNAIL mode, which can degrade the efficiency of parametric processes like conversion and amplification by detuning the system away from optimal conditions. To mitigate these effects, one must run pulses considerably shorter than the maximal Kerr-induced shift $\tau \ll 1/K\overline{n}_{max}$. For reasonable experimental values of $\overline{n}_{max} \sim 1000$ and $K \sim 1\text{KHz}$, this corresponds to $\tau \ll 1\mu\text{sec}$.
Alternative solutions involve chirping the pump tone to dynamically track the Kerr-induced detuning, see for example~Xia et al.\cite{xia_fast_2023}. In addition, cross-Kerr terms can cause back-action into the readout resonator, inadvertently measuring the qubit; these terms limit the amount of photons in the SNAIL, and could impose another limit on the parametric gates and amplification.

\qquad

\emph{Multi-qubit readout and frequency crowding} -- Scalability remains one of the central challenges in quantum-state readout, particularly as the number of qubits per chip increases. In our scheme, multiple readout resonators are coupled to a single embedded SNAIL amplifier. Each qubit can then be selectively addressed through a distinct frequency-conversion process into the SNAIL, enabling sequential multi-qubit readout through a common output channel. This reduces the hardware footprint while preserving coherent on-chip processing. In principle, the same architecture also supports parallel frequency-multiplexed readout, although here we focus on the sequential protocol. However, this multiplexing introduces several key constraints:

First, the number of distinct parametric interactions, such as beam-splitter or two-mode squeezing terms, grows quadratically with the number of coupled modes \cite{mckinney_spectator-aware_2024} (see the Supplementary Material for a detailed discussion). This combinatorial scaling increases the likelihood of accidental resonance: for instance, a pump designed to activate one resonator-SNAIL process might inadvertently drive a qubit-SNAIL swap or a two-photon qubit-resonator transition \cite{naaman_synthesis_2022}. Such unwanted interactions become increasingly probable as the system size grows and resonators are packed more densely in frequency space. Second, parasitic couplings, such as direct leakage between readout resonators and the SNAIL, or between the SNAIL and qubit control ports, can induce residual back-action and degrade fidelity. These effects resemble the well-known static ZZ interactions in multi-qubit architectures, which create crosstalk and frequency shifts that complicate readout and gate operations \cite{ku_suppression_2020,zhao_high-contrast_2020}.

\section{Conclusions}
\label{sec:conclusions}
In this work, we introduced a readout architecture that enables coherent, on-chip processing of quantum-state measurement signals. This is achieved by embedding a nonlinear SNAIL element between the readout and output resonators, allowing for dynamic control over signal flow, amplification, and noise shaping. Our architecture redefines the readout stage as an active, programmable layer within the quantum device—capable of manipulating information before it leaves the cryogenic environment. By integrating amplification, isolation, and signal transformation in a compact and tunable device, the platform eliminates bulky off-chip components and opens a pathway to scalable, frequency-multiplexed quantum readout.

We demonstrated two readout strategies that leverage this coherent processing framework. The first, sequential bosonic gate readout, uses timed parametric interactions to manipulate and route the readout signal with high fidelity and minimal backaction. The second, driven-dissipative readout, realizes directionality through interference and engineered loss, showcasing our platform's ability to implement non-Hamiltonian interactions.

While our results highlight the promise of this approach, practical considerations such as mode leakage, residual dissipation and four-wave mixing, and frequency crowding remain important and would have to be taken into account in the experimental realization. Nonetheless, the ability to coherently manipulate and multiplex readout signals in situ presents a compelling path forward for scalable quantum measurement.

Future directions include extending beyond the quadratic interactions considered in this work to explore richer nonlinear processes for on-chip readout, for example, by deliberately tuning the SNAIL away from its Kerr null-point. 

While quadratic interactions keep the means and covariances separated, nonlinear interactions inherently couple the two. Such interactions could enable more sophisticated information processing directly within the readout chain, that could support more complicated readout processing tasks,including multi-state discrimination, dynamic noise suppression, and adaptive measurement strategies tailored to fluctuating quantum environments \cite{khan_practical_2023, khan_neural_2024}. They also offer a route to mitigating crosstalk, supporting entangled-state discrimination, and implementing error-aware or feedback-enhanced protocols.  Ultimately, the hybridized SNAIL platform paves the way toward a new class of coherent, programmable readout processors that bridge quantum measurement and control at scale.

\emph{Conflict of Interests}. 
Michael Hatridge serves as a consultant for Quantum Circuits, Inc., receiving remuneration in the form of consulting fees, and hold equity in the form of stock options.

\emph{Acknowledgments}. This research was primarily sponsored by the Army Research
Office and was accomplished under Award No. W911NF-23-1-0252 (FastCARS).
The views and conclusions contained in this document are those of the authors and should not be interpreted as representing the official policies, either expressed or implied, of the Army Research Office or the U.S. Government. The U.S. Government is authorized to reproduce and distribute reprints for Government purposes notwithstanding any copyright notation herein. B.M. acknowledges support from the NSF Graduate Research Fellowship Program Fellow No. 2022338601.

\bibliography{references,references3}

@misc{noauthor_lnf-xxxxc4_8b_2025,
	title = {{LNF}-{XXXXC4}\_8B {Datasheet}},
	url = {https://lownoisefactory.com/wp-content/uploads/2025/07/lnf-xxxxc4_8b.pdf},
	publisher = {Low-Noise Factory},
	year = {2025},
}

@article{zhou_realizing_2023,
	title = {Realizing all-to-all couplings among detachable quantum modules using a microwave quantum state router},
	volume = {9},
	copyright = {2023 The Author(s)},
	issn = {2056-6387},
	url = {https://www.nature.com/articles/s41534-023-00723-7},
	doi = {10.1038/s41534-023-00723-7},
	language = {english},
	number = {1},
	urldate = {2023-10-02},
	journal = {npj Quantum Information},
	publisher = {Nature Publishing Group},
	author = {Zhou, Chao and Lu, Pinlei and Praquin, Matthieu and Chien, Tzu-Chiao and Kaufman, Ryan and Cao, Xi and Xia, Mingkang and Mong, Roger S. K. and Pfaff, Wolfgang and Pekker, David and Hatridge, Michael},
	month = jun,
	year = {2023},
	note = {Number: 1},
	pages = {1--9},
}

@article{zhao_high-contrast_2020,
	title = {High-{Contrast} {Z} {Z} {Interaction} {Using} {Superconducting} {Qubits} with {Opposite}-{Sign} {Anharmonicity}},
	volume = {125},
	issn = {0031-9007, 1079-7114},
	url = {https://link.aps.org/doi/10.1103/PhysRevLett.125.200503},
	doi = {10.1103/PhysRevLett.125.200503},
	language = {english},
	number = {20},
	urldate = {2025-06-11},
	journal = {Physical Review Letters},
	author = {Zhao, Peng and Xu, Peng and Lan, Dong and Chu, Ji and Tan, Xinsheng and Yu, Haifeng and Yu, Yang},
	month = nov,
	year = {2020},
	pages = {200503},
}

@article{yurke_su2_1986,
	title = {{SU}(2) and {SU}(1,1) interferometers},
	volume = {33},
	copyright = {http://link.aps.org/licenses/aps-default-license},
	issn = {0556-2791},
	url = {https://link.aps.org/doi/10.1103/PhysRevA.33.4033},
	doi = {10.1103/PhysRevA.33.4033},
	language = {english},
	number = {6},
	urldate = {2025-06-11},
	journal = {Physical Review A},
	author = {Yurke, Bernard and McCall, Samuel L. and Klauder, John R.},
	month = jun,
	year = {1986},
	pages = {4033--4054},
}

@article{weedbrook_gaussian_2012,
	title = {Gaussian quantum information},
	volume = {84},
	copyright = {http://link.aps.org/licenses/aps-default-license},
	issn = {0034-6861, 1539-0756},
	url = {https://link.aps.org/doi/10.1103/RevModPhys.84.621},
	doi = {10.1103/RevModPhys.84.621},
	language = {english},
	number = {2},
	urldate = {2025-06-17},
	journal = {Reviews of Modern Physics},
	author = {Weedbrook, Christian and Pirandola, Stefano and García-Patrón, Raúl and Cerf, Nicolas J. and Ralph, Timothy C. and Shapiro, Jeffrey H. and Lloyd, Seth},
	month = may,
	year = {2012},
	pages = {621--669},
}

@article{sliwa_reconfigurable_2015,
	title = {Reconfigurable {Josephson} {Circulator}/{Directional} {Amplifier}},
	volume = {5},
	copyright = {http://creativecommons.org/licenses/by/3.0/},
	issn = {2160-3308},
	url = {https://link.aps.org/doi/10.1103/PhysRevX.5.041020},
	doi = {10.1103/PhysRevX.5.041020},
	language = {english},
	number = {4},
	urldate = {2025-05-16},
	journal = {Physical Review X},
	author = {Sliwa, K. M. and Hatridge, M. and Narla, A. and Shankar, S. and Frunzio, L. and Schoelkopf, R. J. and Devoret, M. H.},
	month = nov,
	year = {2015},
	pages = {041020},
}

@techreport{schwinger_angular_1952,
	title = {{ON} {ANGULAR} {MOMENTUM}},
	url = {http://www.osti.gov/servlets/purl/4389568-jZqrUJ/native/},
	doi = {10.2172/4389568},
	language = {english},
	number = {NYO-3071, 4389568},
	urldate = {2025-05-29},
	author = {Schwinger, J.},
	month = jan,
	year = {1952},
	pages = {NYO--3071, 4389568},
}

@article{miano_frequency-tunable_2022,
	title = {Frequency-tunable {Kerr}-free three-wave mixing with a gradiometric {SNAIL}},
	volume = {120},
	issn = {0003-6951, 1077-3118},
	url = {https://pubs.aip.org/apl/article/120/18/184002/2833627/Frequency-tunable-Kerr-free-three-wave-mixing-with},
	doi = {10.1063/5.0083350},
	language = {english},
	number = {18},
	urldate = {2025-06-11},
	journal = {Applied Physics Letters},
	author = {Miano, A. and Liu, G. and Sivak, V. V. and Frattini, N. E. and Joshi, V. R. and Dai, W. and Frunzio, L. and Devoret, M. H.},
	month = may,
	year = {2022},
	pages = {184002},
}

@article{ku_suppression_2020,
	title = {Suppression of {Unwanted} {Z} {Z} {Interactions} in a {Hybrid} {Two}-{Qubit} {System}},
	volume = {125},
	issn = {0031-9007, 1079-7114},
	url = {https://link.aps.org/doi/10.1103/PhysRevLett.125.200504},
	doi = {10.1103/PhysRevLett.125.200504},
	language = {english},
	number = {20},
	urldate = {2025-06-11},
	journal = {Physical Review Letters},
	author = {Ku, Jaseung and Xu, Xuexin and Brink, Markus and McKay, David C. and Hertzberg, Jared B. and Ansari, Mohammad H. and Plourde, B. L. T.},
	month = nov,
	year = {2020},
	pages = {200504},
}

@book{helstrom_statistical_1968,
	address = {Oxford New York},
	edition = {2nd ed., rev. and enl},
	series = {International series of monographs in electronics and instrumentation},
	title = {Statistical theory of signal detection},
	isbn = {978-0-08-013265-5},
	language = {english},
	number = {v. 9},
	publisher = {Pergamon Press},
	author = {Helstrom, Carl W.},
	year = {1968},
}

@article{dumas_measurement-induced_2024,
	title = {Measurement-{Induced} {Transmon} {Ionization}},
	volume = {14},
	issn = {2160-3308},
	url = {https://link.aps.org/doi/10.1103/PhysRevX.14.041023},
	doi = {10.1103/PhysRevX.14.041023},
	language = {english},
	number = {4},
	urldate = {2025-05-16},
	journal = {Physical Review X},
	author = {Dumas, Marie Frédérique and Groleau-Paré, Benjamin and McDonald, Alexander and Muñoz-Arias, Manuel H. and Lledó, Cristóbal and D’Anjou, Benjamin and Blais, Alexandre},
	month = oct,
	year = {2024},
	pages = {041023},
}

@book{chen_linear_1999,
	address = {New York},
	edition = {3rd ed},
	series = {The {Oxford} series in electrical and computer engineering},
	title = {Linear system theory and design},
	isbn = {978-0-19-511777-6 978-1-61344-115-2 978-0-19-511778-3},
	language = {english},
	publisher = {Oxford University Press},
	editor = {Chen, Chi-Tsong},
	year = {1999},
}

@article{bello_systematic_2025,
	title = {Systematic time-coarse-graining for driven quantum systems},
	volume = {23},
	issn = {2331-7019},
	url = {https://link.aps.org/doi/10.1103/PhysRevApplied.23.054042},
	doi = {10.1103/PhysRevApplied.23.054042},
	language = {English},
	number = {5},
	urldate = {2025-06-11},
	journal = {Physical Review Applied},
	author = {Bello, Leon and Fan, Wentao and Gandotra, Aditya and Türeci, Hakan E.},
	month = may,
	year = {2025},
	pages = {054042},
}

@book{wiseman_quantum_2009,
	edition = {1},
	title = {Quantum {Measurement} and {Control}},
	copyright = {https://www.cambridge.org/core/terms},
	isbn = {978-0-521-80442-4 978-0-511-81394-8 978-1-107-42415-9},
	url = {https://www.cambridge.org/core/product/identifier/9780511813948/type/book},
	doi = {10.1017/CBO9780511813948},
	urldate = {2025-12-19},
	publisher = {Cambridge University Press},
	author = {Wiseman, Howard M. and Milburn, Gerard J.},
	month = nov,
	year = {2009},
}

@misc{xia_exceeding_2025,
	title = {Exceeding the {Parametric} {Drive} {Strength} {Threshold} in {Nonlinear} {Circuits}},
	copyright = {Creative Commons Attribution 4.0 International},
	url = {https://arxiv.org/abs/2506.03456},
	doi = {10.48550/ARXIV.2506.03456},
	urldate = {2025-06-06},
	publisher = {arXiv},
	author = {Xia, Mingkang and Lledó, Cristóbal and Capocci, Matthew and Repicky, Jacob and D'Anjou, Benjamin and Mondragon-Shem, Ian and Kaufman, Ryan and Koch, Jens and Blais, Alexandre and Hatridge, Michael},
	year = {2025},
	note = {Version Number: 1},
}

@misc{khan_neural_2024,
	title = {A neural processing approach to quantum state discrimination},
	copyright = {Creative Commons Attribution 4.0 International},
	url = {https://arxiv.org/abs/2409.03748},
	doi = {10.48550/ARXIV.2409.03748},
	urldate = {2025-06-30},
	publisher = {arXiv},
	author = {Khan, Saeed A. and Hu, Fangjun and Angelatos, Gerasimos and Hatridge, Michael and Türeci, Hakan E.},
	year = {2024},
	note = {Version Number: 2},
}

@misc{mckinney_spectator-aware_2024,
	title = {Spectator-{Aware} {Frequency} {Allocation} in {Tunable}-{Coupler} {Quantum} {Architectures}},
	copyright = {arXiv.org perpetual, non-exclusive license},
	url = {https://arxiv.org/abs/2409.18262},
	doi = {10.48550/ARXIV.2409.18262},
	urldate = {2025-06-20},
	publisher = {arXiv},
	author = {McKinney, Evan and Yusuf, Israa G. and Falstin, Girgis and Agarwal, Gaurav and Hatridge, Michael and Jones, Alex K.},
	year = {2024},
	note = {Version Number: 4},
}

@article{friis_noise_1944,
	title = {Noise {Figures} of {Radio} {Receivers}},
	volume = {32},
	copyright = {https://ieeexplore.ieee.org/Xplorehelp/downloads/license-information/IEEE.html},
	issn = {0096-8390},
	url = {http://ieeexplore.ieee.org/document/1695024/},
	doi = {10.1109/JRPROC.1944.232049},
	number = {7},
	urldate = {2025-02-05},
	journal = {Proceedings of the IRE},
	author = {Friis, H.T.},
	month = jul,
	year = {1944},
	pages = {419--422},
}

@article{macklin_nearquantum-limited_2015,
	title = {A near–quantum-limited {Josephson} traveling-wave parametric amplifier},
	volume = {350},
	url = {https://www.science.org/doi/10.1126/science.aaa8525},
	doi = {10.1126/science.aaa8525},
	number = {6258},
	urldate = {2025-03-27},
	journal = {Science},
	author = {Macklin, C. and O’Brien, K. and Hover, D. and Schwartz, M. E. and Bolkhovsky, V. and Zhang, X. and Oliver, W. D. and Siddiqi, I.},
	month = oct,
	year = {2015},
	note = {Publisher: American Association for the Advancement of Science},
	pages = {307--310},
}

@book{pozar_microwave_2012,
	edition = {4th},
	title = {Microwave {Engineering}},
	publisher = {Wiley},
	author = {Pozar, David M.},
	year = {2012},
}

@misc{malnou_traveling-wave_2024,
	title = {A {Traveling}-{Wave} {Parametric} {Amplifier} and {Converter}},
	url = {http://arxiv.org/abs/2406.19476},
	doi = {10.48550/arXiv.2406.19476},
	urldate = {2025-03-19},
	publisher = {arXiv},
	author = {Malnou, M. and Miller, B. T. and Estrada, J. A. and Genter, K. and Cicak, K. and Teufel, J. D. and Aumentado, J. and Lecocq, F.},
	month = jun,
	year = {2024},
	note = {arXiv:2406.19476 [quant-ph]},
}

@misc{metelmann_quantum-limited_2022,
	title = {Quantum-limited amplification without instability},
	url = {http://arxiv.org/abs/2208.00024},
	doi = {10.48550/arXiv.2208.00024},
	urldate = {2025-02-28},
	publisher = {arXiv},
	author = {Metelmann, A. and Lanes, O. and Chien, T.-Z. and McDonald, A. and Hatridge, M. and Clerk, A. A.},
	month = jul,
	year = {2022},
	note = {arXiv:2208.00024 [quant-ph]},
}

@article{gard_fast_2024,
	title = {Fast high-fidelity quantum nondemolition readout of a superconducting qubit with tunable transverse couplings},
	volume = {21},
	url = {https://link.aps.org/doi/10.1103/PhysRevApplied.21.024008},
	doi = {10.1103/PhysRevApplied.21.024008},
	number = {2},
	urldate = {2024-08-05},
	journal = {Physical Review Applied},
	author = {Gard, Bryan T. and Parrott, Zachary and Jacobs, Kurt and Aumentado, José and Simmonds, Raymond W.},
	month = feb,
	year = {2024},
	note = {Publisher: American Physical Society},
	pages = {024008},
}

@article{khezri_measurement-induced_2023,
	title = {Measurement-{Induced} {State} {Transitions} in a {Superconducting} {Qubit}: {Within} the {Rotating} {Wave} {Approximation}},
	volume = {20},
	issn = {2331-7019},
	shorttitle = {Measurement-{Induced} {State} {Transitions} in a {Superconducting} {Qubit}},
	url = {http://arxiv.org/abs/2212.05097},
	doi = {10.1103/PhysRevApplied.20.054008},
	number = {5},
	urldate = {2024-06-18},
	journal = {Physical Review Applied},
	author = {Khezri, Mostafa and Opremcak, Alex and Chen, Zijun and Miao, Kevin C. and McEwen, Matt and Bengtsson, Andreas and White, Theodore and Naaman, Ofer and Sank, Daniel and Korotkov, Alexander N. and Chen, Yu and Smelyanskiy, Vadim},
	month = nov,
	year = {2023},
	note = {arXiv:2212.05097 [quant-ph]},
	pages = {054008},
}

@article{clerk_introduction_2010,
	title = {Introduction to quantum noise, measurement, and amplification},
	volume = {82},
	url = {https://link.aps.org/doi/10.1103/RevModPhys.82.1155},
	doi = {10.1103/RevModPhys.82.1155},
	number = {2},
	urldate = {2024-05-14},
	journal = {Reviews of Modern Physics},
	author = {Clerk, A. A. and Devoret, M. H. and Girvin, S. M. and Marquardt, Florian and Schoelkopf, R. J.},
	month = apr,
	year = {2010},
	note = {Publisher: American Physical Society},
	pages = {1155--1208},
}

@article{caves_quantum_1982,
	title = {Quantum limits on noise in linear amplifiers},
	volume = {26},
	url = {https://link.aps.org/doi/10.1103/PhysRevD.26.1817},
	doi = {10.1103/PhysRevD.26.1817},
	number = {8},
	urldate = {2024-02-19},
	journal = {Physical Review D},
	author = {Caves, Carlton M.},
	month = oct,
	year = {1982},
	note = {Publisher: American Physical Society},
	pages = {1817--1839},
}

@misc{xia_fast_2023,
	title = {Fast superconducting qubit control with sub-harmonic drives},
	url = {http://arxiv.org/abs/2306.10162},
	doi = {10.48550/arXiv.2306.10162},
	urldate = {2024-01-25},
	publisher = {arXiv},
	author = {Xia, Mingkang and Zhou, Chao and Liu, Chenxu and Patel, Param and Cao, Xi and Lu, Pinlei and Mesits, Boris and Mucci, Maria and Gorski, David and Pekker, David and Hatridge, Michael},
	month = jun,
	year = {2023},
	note = {arXiv:2306.10162 [quant-ph]},
}

@misc{khan_practical_2023,
	title = {Practical trainable temporal post-processor for multi-state quantum measurement},
	url = {http://arxiv.org/abs/2310.18519},
	doi = {10.48550/arXiv.2310.18519},
	urldate = {2024-01-04},
	publisher = {arXiv},
	author = {Khan, Saeed A. and Kaufman, Ryan and Mesits, Boris and Hatridge, Michael and Türeci, Hakan E.},
	month = oct,
	year = {2023},
	note = {arXiv:2310.18519 [quant-ph]},
}

@article{lecocq_nonreciprocal_2017,
	title = {Nonreciprocal {Microwave} {Signal} {Processing} with a {Field}-{Programmable} {Josephson} {Amplifier}},
	volume = {7},
	url = {https://link.aps.org/doi/10.1103/PhysRevApplied.7.024028},
	doi = {10.1103/PhysRevApplied.7.024028},
	number = {2},
	urldate = {2023-11-28},
	journal = {Physical Review Applied},
	author = {Lecocq, F. and Ranzani, L. and Peterson, G. A. and Cicak, K. and Simmonds, R. W. and Teufel, J. D. and Aumentado, J.},
	month = feb,
	year = {2017},
	note = {Publisher: American Physical Society},
	pages = {024028},
}

@article{naaman_synthesis_2022,
	title = {Synthesis of {Parametrically} {Coupled} {Networks}},
	volume = {3},
	url = {https://link.aps.org/doi/10.1103/PRXQuantum.3.020201},
	doi = {10.1103/PRXQuantum.3.020201},
	number = {2},
	urldate = {2022-12-12},
	journal = {PRX Quantum},
	author = {Naaman, Ofer and Aumentado, José},
	month = may,
	year = {2022},
	note = {Publisher: American Physical Society},
	pages = {020201},
}

@article{blais_cavity_2004,
	title = {Cavity quantum electrodynamics for superconducting electrical circuits: {An} architecture for quantum computation},
	volume = {69},
	shorttitle = {Cavity quantum electrodynamics for superconducting electrical circuits},
	url = {https://link.aps.org/doi/10.1103/PhysRevA.69.062320},
	doi = {10.1103/PhysRevA.69.062320},
	number = {6},
	urldate = {2022-07-14},
	journal = {Physical Review A},
	author = {Blais, Alexandre and Huang, Ren-Shou and Wallraff, Andreas and Girvin, S. M. and Schoelkopf, R. J.},
	month = jun,
	year = {2004},
	note = {Publisher: American Physical Society},
	pages = {062320},
}

@article{rosenthal_efficient_2021,
	title = {Efficient and {Low}-{Backaction} {Quantum} {Measurement} {Using} a {Chip}-{Scale} {Detector}},
	volume = {126},
	url = {https://link.aps.org/doi/10.1103/PhysRevLett.126.090503},
	doi = {10.1103/PhysRevLett.126.090503},
	number = {9},
	urldate = {2022-02-08},
	journal = {Physical Review Letters},
	author = {Rosenthal, Eric I. and Schneider, Christian M. F. and Malnou, Maxime and Zhao, Ziyi and Leditzky, Felix and Chapman, Benjamin J. and Wustmann, Waltraut and Ma, Xizheng and Palken, Daniel A. and Zanner, Maximilian F. and Vale, Leila R. and Hilton, Gene C. and Gao, Jiansong and Smith, Graeme and Kirchmair, Gerhard and Lehnert, K. W.},
	month = mar,
	year = {2021},
	note = {Publisher: American Physical Society},
	pages = {090503},
}

@article{blais_circuit_2021,
	title = {Circuit quantum electrodynamics},
	volume = {93},
	url = {https://link.aps.org/doi/10.1103/RevModPhys.93.025005},
	doi = {10.1103/RevModPhys.93.025005},
	number = {2},
	urldate = {2022-02-03},
	journal = {Reviews of Modern Physics},
	author = {Blais, Alexandre and Grimsmo, Arne L. and Girvin, S. M. and Wallraff, Andreas},
	month = may,
	year = {2021},
	note = {Publisher: American Physical Society},
	pages = {025005},
}

\appendix

\section{Derivation of the effective Hamiltonian}
\label{app:effective_Hamiltonian_derivation}
The basic EA building-block is comprised of a qubit-resonator system and an output mode, coupled to a common highly nonlinear SNAIL mode. Although the platform can be extended to multiple qubit-resonator systems, in this derivation we focus on a single readout unit. The derivation can be easily generalized to multiple qubit-resonator systems in the same way. 

We start with the raw Hamiltonian of the readout unit in the lab frame, given by the sum of the following Hamiltonians,
\begin{subequations}
    \begin{align}
    &\hat{H}_0 = \omega_a \hat{a}^\dagger \hat{a} + \omega_b \hat{b}^\dagger \hat{b} + \omega_s \hat{s}^\dagger \hat{s}  \\
    &\hat{H}_{qb} = \hat{\sigma}_z \hat{a}^\dagger \hat{a} =\pm \chi \hat{a}^\dagger \hat{a}\\
    &\hat{H}_L = g_{as} \left (\hat{a}^\dagger \hat{s} + \hat{s}^\dagger \hat{a} \right ) + g_{bs} \left (\hat{b}^\dagger \hat{s} + \hat{s}^\dagger \hat{b} \right )\\
    &\hat{H}_{NL} = g_3(\hat{s} + \hat{s}^\dagger)^3 \\
    &\hat{H}_{d} = (\eta^*(t)\hat{a} + \eta(t)\hat{a}^\dagger) + (\epsilon^*(t) \hat{s} + \epsilon(t) \hat{s}^\dagger)
\end{align}
\end{subequations}
Note that we assume operation in the dispersive regime, and that the qubit is in a well defined computational state.

The heart of the method depends on the linear coupling described by the interaction in $\hat{H}_L$, coupling the readout and output resonators to the SNAIL. Since the frequency separation between the resonators is quite large compared to the strength of the couplings, this coupling is generally weak -- but \emph{not} negligible. 

To make this explicit, we focus on the $\hat{H}_0 + \hat{H}_L$ part of the Hamiltonian, which we can write as a simple quadratic form.
\begin{equation}
    \hat{H}_0 + \hat{H}_L = \vec{a}^\dagger(\mathbf{\Omega} + \mathbf{V}) \vec{a}
\end{equation}
where $\hat{a} = (\hat{a}, \hat{s}, \hat{b})^\intercal$, and we define the coupling matrices,
\begin{equation}
    \mathbf{\Omega}_0 = \begin{pmatrix}
		\omega_a & 0 & 0 \\
		0 & \omega_b & 0 \\
		0 & 0 & \omega_c
	\end{pmatrix} \qquad\mathbf{V} = \begin{pmatrix}
		0 & 0 & g_{as} \\
		0 & 0 & g_{bs} \\
		g_{as}^* & g_{bs}^* & 0 \end{pmatrix}
\end{equation}
Although this can be diagonalized and solved exactly, the eigenvalues of the full matrix are the roots of a qubic polynomial which is algebraically complex and difficult to work with. Instead, since we know that the coupling is weak compared 
to the frequency separations in the system, we can apply non-degenerate perturbation theory to systematically diagonalize the system and find the corrections due to the coupling.

We expand the operators and frequencies up to second-order in the couplings $g_{as}, g_{bs}$, written here explicitly.

For the \emph{readout resonator},
\begin{align}
\hat{\alpha}^{(0)} &= \hat{a} 
    & \quad \omega_a^{(0)} &= \omega_a \\
\hat{\alpha}^{(1)} &= -\frac{g_{as}}{\Delta_{as}} \hat{s} 
    & \quad \omega_a^{(1)} &= 0 \\
\hat{\alpha}^{(2)} &= -\frac{g_{bs} g_{as}}{\Delta_{ab} \Delta_{as}} \hat{b} 
    & \quad \omega_a^{(2)} &= \frac{|g_{as}|^2}{\Delta_{as}}
\end{align}
where $\Delta_{as} = \omega_a - \omega_s $. For the inverse transformation,
\begin{subequations}
\begin{align}
	&\hat{a}^{(0)} = \hat{\alpha} \\
	&\hat{a}^{(1)} = \frac{g_{as}}{\Delta_{as}} \hat{\gamma}  \\
	&\hat{a}^{(2)} = \frac{g_{bs} g_{as}}{\Delta_{ab} \Delta_{as}} \hat{\beta} 
\end{align}
\end{subequations}

\begin{center}
\rule{0.5\linewidth}{0.4pt}
\end{center}

For the \emph{output resonator}, 
\begin{align}
\hat{\beta}^{(0)} &= \hat{b} 
    & \quad \omega_b^{(0)} &= \omega_b \\
\hat{\beta}^{(1)} &= -\frac{g_{bs}}{\Delta_{bs}} \hat{s} 
    & \quad \omega_b^{(1)} &= 0 \\
\hat{\beta}^{(2)} &= -\frac{g_{bs} g_{as}}{\Delta_{ab} \Delta_{bs}} \hat{a} 
    & \quad \omega_b^{(2)} &= \frac{|g_{as}|^2}{\Delta_{as}}
\end{align}
where $\Delta_{bs} = \omega_{b} - \omega_s$. For the inverse transformation,
\begin{subequations}
    \begin{align}
	&\hat{b}^{(0)} = \hat{\beta} \\
	&\hat{b}^{(1)} = \frac{g_{bs}}{\Delta_{bs}} \hat{\gamma}  \\
	&\hat{b}^{(2)} = \frac{g_{bs} g_{as}}{\Delta_{bs} \Delta_{as}} \hat{\alpha} 
\end{align}
\end{subequations}

\begin{center}
\rule{0.5\linewidth}{0.4pt}
\end{center}

For the SNAIL resonator, 
\begin{align}
\hat{\gamma}^{(0)} &= \hat{s}  
    & \quad \omega_s^{(0)} &= \omega_s \\
\hat{\gamma}^{(1)} &= -\frac{g_{as}^*}{\Delta_{as}} \hat{a} - \frac{g_{bs}^*}{\Delta_{bs}} \hat{b} 
    & \quad \omega_s^{(1)} &= 0 \\
\hat{\gamma}^{(2)} &= -\frac{g_{as}^* g_{bs}^*}{\Delta_{as} \Delta_{bs}} \left( \hat{a} + \hat{b} \right) 
    & \quad \omega_s^{(2)} &=  -\frac{|g_{as}|^2}{\Delta_{as}} - \frac{|g_{bs}|^2}{\Delta_{bs}}
\end{align}
For the inverse transformation,
\begin{subequations}
    \begin{align}
	&\hat{s}^{(0)} = \hat{\gamma} \\
	&\hat{s}^{(1)} = \frac{g_{as}^*}{\Delta_{as}} \hat{\alpha} + \frac{g_{bs}^*}{\Delta_{bs}} \hat{\beta}  \\
	&\hat{s}^{(2)} = \frac{g_{bs}^* g_{as}^*}{\Delta_{bs} \Delta_{as}} (\hat{\alpha} +\hat{\beta)}
\end{align}
\end{subequations}

\begin{center}
\rule{0.5\linewidth}{0.4pt}
\end{center}

Note that up to first-order in the couplings, there is no change to the frequencies, and the resonators are weakly hybridized only with the SNAIL, to which they are directly coupled. Going up to second-order in the couplings we observe two effects: a small dispersive shift to all modes, directly proportional to the strength of their hybridization with the SNAIL resonator; it reveals direct coupling and hybridization between the readout and output resonator through the SNAIL.

The form of the corrections motivates us to define the \emph{hybridization strengths},
\begin{equation}
    \lambda_{ij} \equiv \frac{g_{ij}}{|\Delta_{ij}|} \qquad i \in\{a,b\}.
\end{equation}
Since the detuning between the resonators is large, this provides a better parameter for the problem. We now plug that into the Hamiltonian, and only keep terms to first order in the hybridization strength, rather than the couplings.
\begin{subequations}
    \begin{align}
        \hat{H}_0 &\approx \omega_a \hat{\alpha}^\dagger \hat{\alpha} + \omega_b \hat{\beta}^\dagger \hat{\beta} + \omega_s \hat{\gamma}^\dagger \hat{\gamma}  \\
        \hat{H}_{qb} &\approx \chi \hat{\alpha}^\dagger \hat{\alpha} \\
        \hat{H}_{NL} &\approx g_3(\hat{\gamma} - \lambda_{as} \hat{\alpha} - \lambda_{bs} \hat{\beta} + h.c.)^3
    \end{align}
\end{subequations}

It is also useful to look at the Hamiltonian expanded to second-order in the hybridization strength,
\begin{subequations}
\begin{align}
    \hat{H}_0 &\approx \omega_\alpha \hat{\alpha}^\dagger \hat{\alpha} + \omega_\beta \hat{\beta}^\dagger \hat{\beta} + \omega_\gamma \hat{\gamma}^\dagger \hat{\gamma} \\
	\hat{H}_{qb} &\approx \chi \hat{\alpha}^\dagger \hat{\alpha} + \chi(\lambda_{as} \hat{\alpha}^\dagger \hat{\gamma} + \lambda_{as}^* \hat{\gamma}^\dagger \hat{\alpha}) \\
    \begin{split}
    \hat{H}_{NL} &\approx g_3 \big( \hat{\gamma} - (\lambda_{as} + \lambda_{as}\lambda_{bs})\hat{\alpha}  \\ 
   \qquad - &(\lambda_{bs} + \lambda_{bs}\lambda_{as})  \hat{\beta} + h.c. \big)^3
    \end{split}
\end{align}
\end{subequations}
Note that the second term in $\hat{H}_{qb}$ is second-order in the hybridization, assuming $\chi \ll \Delta_{as}$. This term indicates a hybridization between the SNAIL and the qubit; in other words, the SNAIL photons "measure" the qubit, and thus limit the number of photons we can put in the SNAIL.

Additionally, because of the hybridization the drive channels of one resonator couple also to the others.
\begin{equation}
\begin{split}
	\hat{H}_{d} &= (\eta^*(t)\hat{\alpha} + \eta(t)\hat{\alpha}^\dagger) \\
	& + \left(\lambda_{as} \eta^*(t) \hat{\gamma} + \lambda_{as}^* \eta(t) \hat{\gamma}^\dagger \right) \\
	& +\left(\lambda_{as} \lambda_{ab} \eta^*(t) \hat{\beta} + \lambda_{as}^* \lambda_{ab}^* \eta(t) \hat{\beta}^\dagger \right)\\
	& + \left(\epsilon^*(t) \hat{\gamma} + \epsilon(t) \hat{\gamma}^\dagger \right) \\
	& + \left((\lambda_{as} + \frac{\Delta_{as}}{\Delta_{ab}}\lambda_{as} \lambda_{bs}) \epsilon^*(t) \hat{\alpha} +  h.c.\right) \\
	& + \left((\lambda_{bs} + \frac{\Delta_{bs}}{\Delta_{ab}}\lambda_{as} \lambda_{bs}) \epsilon^*(t) \hat{\beta} + h.c. \right)
\end{split}  
\end{equation}
These terms describe the drives "leaking" from one resonator another.
Depending on the frequency of the drive, these terms would be non-resonant and effectively small under standard operation. 

To obtain the effective Hamiltonian, we will transform into the frame of $\hat{H}_0$ and take the RWA.
\begin{subequations}
    \begin{align}
	&\hat{H}_{qb} \approx \chi \hat{\alpha}^\dagger \hat{\alpha} \\
	&\hat{H}_{NL} \approx g_{3}(\hat{\gamma}e^{-i\omega_s t} - \lambda_{as} \hat{\alpha}e^{-i\omega_a t} - \lambda_{bs}\hat{\beta} e^{-i\omega_b t} + h.c.)^3
\end{align}
\end{subequations}
In order to obtain the form in the main paper, we need to take another step. We assume that the SNAIL is driven by a strong pump field at frequency $\omega_p$ and amplitude $\epsilon_p$, and that this pump is generally highly detuned from the SNAIL resonance. In this regime of operations, we can assume that its dynamics can be adiabatically eliminated, and the SNAIL amplitude simply follows the drive.
\begin{equation}
    \hat{\gamma}' = \hat{\gamma} + p_0 e^{-i\omega_p t}
\end{equation}
where $p_0 = |\epsilon|/\Delta_{sp}$ is determined by the steady-state of the SNAIL, determined by pump detuning and amplitude. 

Expanding $\hat{H}_{NL}$ and keeping only the resonant terms in the RWA, gives us the expression for the effective Hamiltonian from the main text,
\begin{subequations}
\begin{align*}
    &\hat{H}_{\text{eff}} = g_{\omega_p \omega_i \omega_j} \ p_{\omega_p} \hat{a}_{\omega_i} \hat{a}_{\omega_j} e^{-i\Delta_{\omega_p \omega_i \omega_j} t} + h.c., \\ 
    &\text{s.t.} \quad \omega_p + \omega_i + \omega_j \equiv \Delta_{\omega_p \omega_i \omega_j} \ll \min(\omega_p,\omega_i, \omega_j).
\end{align*}
\end{subequations}

\begin{center}
\rule{0.5\linewidth}{0.4pt}
\end{center}

We reiterate the relevant regime of operation:
\begin{enumerate}
    \item The linear couplings and the qubit dispersive shift must be weak compared to the frequency separation between the resonators $g_{as}, g_{bs} \ll \Delta_{as}, \Delta_{bs}$.
    \item  The SNAIL is driven far off-resonance, such that its off-resonance population can be adiabatically eliminated.
    \item The three-wave mixing nonlinearity of the SNAIL is weak compared to the frequency separations in the system $g_3 \ll \Delta_{as}, \Delta_{bs}$.
    \item The pump strength and the induced couplings are weak compared to the resonator frequency differences.
In this regime, the system acts linearly, with frequency multiplexed quadratic interactions.
\end{enumerate}

\subsection{Drive rates and effective couplings}
The effective coupling rate for a process involving modes at frequencies $\omega_{i (j)}$ is related to the pump-induced displacement of the SNAIL, which is in turn related to the pump rate into the SNAIL.

As discussed above, the effective coupling rates are given by,
\begin{equation}
    g_{\omega_i\omega_j} = \lambda_{is} \lambda_{js} g_3 \cdot p_{-\omega_i-\omega_j}
\end{equation}
where $p_{\omega}$ is the SNAIL displacement value at frequency $\omega$, and the hybridization strengths are defined as above.

The displacement value is assumed to be adiabatically determined by the SNAIL drive amplitude $\epsilon(t)$. Denoting $\sigma = \langle \hat{s} \rangle$ and writing the equation of motion in the drive frame,
\begin{equation}
    \dot{\sigma} = -(i\Delta_{sp} + \gamma_s )\sigma + \epsilon_{\omega_p}(t)
\end{equation}
where $\epsilon_{\omega_p}$ is the amplitude of the drive at frequency $\omega_p$. In steady-state, denoting the steady state value by $p$,
\begin{equation}
    p_{\omega_p} = \frac{\epsilon_{\omega_p}}{\gamma_s + i \Delta_{sp}} \approx -\frac{i\epsilon_{\omega_p}}{\Delta_{sp}}
\end{equation}
where we assumed that $\Delta_{sp} \gg \gamma_s$, which is the case in the experiment. Plugging that in,
\begin{equation}
    g_{\omega_i \omega_j} = -i \left( \lambda_{is} \lambda_{js} \frac{g_3}{\Delta_{sp}} \right) \epsilon_{-\omega_i-\omega_j}
\end{equation}

Or alternatively,
\begin{equation}
    |\epsilon_{-\omega_i - \omega_j}| = \frac{g_{\omega_i \omega_j}}{g_3} \cdot \frac{|\omega_s - \omega_p|}{\lambda_{is}\lambda_{js}}
\end{equation}
Assuming reasonable experimental values, of $g_3 \approx 30\text{MHz}$ and hybridizations of about $0.1$, we find that for coupling strength of a few MHz, very feasible drive rates of $1-2\text{GHz}$ would be required.

\section{Derivation of the equations for the means and covariance matrix}
\label{app:cov_eoms}
A closed linear quantum system can be efficiently described using symplectic transformations \cite{weedbrook_gaussian_2012} on the quadrature mode operators $\vec{X} = (\hat{X}_1, \hat{P}_1, \cdots, \hat{X}_n, \hat{P}_n)^T$. In the case where the input state is Gaussian, the system and its dynamics are completely determined by the means and covariance matrix.

We define a linear system as one described by a quadratic Hamiltonian,
\begin{equation}
    \hat{H} = \vec{X}^T \mathbf{H} \vec{X} = \sum_{ij} H_{ij} \hat{X}_i \hat{X}_j
\end{equation}
where $\mathbf{H}$ is a hermitian matrix of coupling and $\vec{X}=(\hat{q}_1, \hat{p}_1, \ldots, \hat{q}_n, \hat{p}_n)^\intercal$. 

Although not common, a similarly efficient description remains correct also when the resonator decay is added into the system, which we derive here. 

In the following, we show that the dynamics are generated by a new matrix,
\begin{subequations}
\begin{align}
	&\mathbf{G} = \mathbf{\Omega H} \\
	&[\mathbf{G}]_{ij} = -i\frac{d}{d\langle \hat{X}_j\rangle} \langle [\hat{X}_i, \hat{H}]\rangle 
\end{align} 
\end{subequations}
where $\mathbf{\Omega} = \bigoplus_{i=1}^{2n} \mathbf{\Omega}_2$ is the block diagonal symplectic form matrix, capturing the commutation relations of $\vec{X}$.

The matrix $\mathbf{G}$ satisfies $\Omega$-skew-symmetry,
\begin{equation}
    	\mathbf{G}^\intercal\mathbf{\Omega} + \mathbf{\Omega}\mathbf{G} = 0
\end{equation}
which indicates that the exponential map $\exp(\mathbf{G}t)$ is symplectic. Each block in $\mathbf{G}$ must satisfy this constraint. For a basic $2\times 2$ block,
\begin{equation}
\mathbf{G}_{ij} = \begin{pmatrix}
		A & B \\ C & -A
	\end{pmatrix}
\end{equation}

Our starting point would be the Markov-Lindblad approximation, which gives the operator equations of motion,
\begin{equation}
    \hat{X}_i = -i[\hat{X}_i, \hat{H}] + \sum_j \gamma_j \mathcal{D}_j[\hat{X}_i]
\end{equation}
where $\mathcal{D}_i[\hat{o}] = \hat{a}_i^\dagger \hat{o} \hat{a}_i - \frac{1}{2}[\hat{a}_i \hat{a}_i^\dagger, \hat{o}]_+$ is the Lindblad superoperator associated with resonator decay.

\subsection{Dynamics of the means}
For the means, the Lindblad equations become,
\begin{equation}
    	\langle \hat{X}_i \rangle = -i\langle[\hat{X}_i, \hat{H}]\rangle + \sum_j \gamma_j \langle \mathcal{D}_j[\hat{X}_i]\rangle
\end{equation}
we will use this as our starting point for the derivation.

\subsubsection{Hamiltonian contribution}
For linear systems (i.e. Hamiltonians with only quadratic terms), the commutator term generates a linear combination of the quadrature operators,
\begin{equation}
    \frac{d}{dt} \langle \hat{X}_i \rangle_H = -i\langle[\hat{X}_i, \hat{H}]\rangle = \sum_j {G}_{ij} \langle \hat{X}_j \rangle_H + C_i
\end{equation}
where we defined,
\begin{subequations}
    \begin{align}
	&{G}_{ij} = -i\frac{d}{d\langle \hat{X}_j\rangle} \langle [\hat{X}_i, \hat{H}]\rangle \\
	&C_i = -i\langle[ \hat{X}_i, \hat{H}]\rangle_{\bigg |_{\langle \vec{X}\rangle=0}}
\end{align}
\end{subequations}
where $G_{ij}$ relates the dynamics to other operators, and $C_i$ are constant terms (for example, due to a drive). The $H$ subscript denotes that this is the evolution without the dissipative terms.
\begin{align*}
	G_{ij} &= -i \frac{d}{d\langle \hat{X}_j\rangle} \sum_{kl} H_{kl} [ \hat{X_i},  \hat{X}_k \hat{X}_l]  \\ 
    &= -i \frac{d}{d\langle \hat{X}_j\rangle}\sum_{kl} H_{kl} \left ( [\hat{X}_i, \hat{X}_k] \hat{X}_l + \hat{X}_k [\hat{X}_i, \hat{X}_l] \right ) \\ 
	&= -i \frac{d}{d\langle \hat{X}_j\rangle}\sum_{kl} H_{kl} \left( \underbrace{[\hat{X}_i, \hat{X}_l]}_{\frac{i}{2} \Omega_{il}} + \underbrace{[\hat{X}_i, \hat{X}_k]}_{\frac{i}{2} \Omega_{ik}} \right) \\ 
    &= \sum_{k} \Omega_{ik} H_{kj}  =[\mathbf{\Omega\mathbf{H}}]_{ij}
\end{align*}
\subsubsection{Dissipative contribution}
Resonator decay keeps the system closed, and again only couples the quadrature operator to other quadrature operators. Even more so, it does not couple between different quadrature amplitudes,
\begin{equation}
    \langle \hat{X}_i \rangle_D = \gamma_j \langle \mathcal{D}_j[\hat{X}_i] \rangle = -\frac{1}{2}\gamma_i \delta_{ij} \langle \hat{X}_j \rangle
\end{equation}
which defines another, diagonal matrix,
\begin{equation}
    \mathbf{\Gamma}_{ij} = \gamma_i \delta_{ij}
\end{equation}

\begin{center}
\rule{0.5\linewidth}{0.4pt}
\end{center}

Overall, we can write the evolution of the means in a linear open quantum system in matrix form,
\begin{equation}
    \frac{d}{dt}\vec{\mu} = \left(\mathbf{\Omega H} - \frac{\mathbf{\Gamma}}{2} \right) \vec{\mu} + \vec{C}
\end{equation}
where we defined $\vec{\mu} \equiv \langle \vec{X} \rangle$, which differs from the closed system case by the diagonal matrix $\mathbf{\Gamma}$, breaking the symplectic structure and introducing decay.

\subsection{Dynamics of the covariance matrix}
To obtain a complete description of the system in the linear case, we have to also calculate the covariance matrix. Due to the non-commutativity of the quadrature amplitudes, the covariance matrix is defined as,
\begin{equation}
    	{V}_{ij} = \frac{1}{2} \left( \langle \hat{X}_i \hat{X}_j \rangle + \langle \hat{X}_j \hat{X}_i \rangle\right) - \langle \hat{X}_j  \rangle \langle \hat{X}_i \rangle
\end{equation}
In the case of a closed system, the evolution of the covariance matrix is simply given by $\mathbf{U}(t)\mathbf{V}\mathbf{U}^T(t)$, so no different description is necessary. However, in the open systems case, the time evolution of the second-moments depends also on the dissipative contributions,
\begin{equation}
    	\frac{d}{dt} \langle \hat{X}_i \hat{X}_j \rangle = -i \langle [\hat{X}_i \hat{X}_j, \hat{H}] \rangle + \sum_i \gamma_i\langle \mathcal{D}_i[\hat{X}_i \hat{X}_j]\rangle
\end{equation}
As we will see, simple resonator decay keeps the equations of motion  closed, which will allow us to write the dynamics as a simple matrix differential equation.
\begin{align*}
	\frac{d}{dt} V_{ij} &=  
	\frac{d}{dt} \left[\frac{1}{2} \left( \langle \hat{X}_i \hat{X}_j \rangle + \langle \hat{X}_j \hat{X}_i \rangle\right) - \langle \hat{X}_j  \rangle \langle \hat{X}_i \rangle \right] \\
	&= \sum_{kl} G_{ij}^{kl} {V}_{kl} - \sum_{kl} \frac{1}{2}\Gamma_{ij}^{kl} {V}_{kl} + C_{ij} \\
    &= \sum_{kl} \left(G_{ij}^{kl} - \frac{1}{2} \Gamma_{ij}^{kl} \right){V}_{kl} + C_{ij} - \frac{1}{2}D_{ij}\\
\end{align*}
where we define the Hamiltonian tensors,
\begin{align*}
	G_{ij}^{kl} &= -i\frac{d}{d {V}_{kl}} \bigg( \frac{1}{2} \left( \langle[\hat{X}_{i} \hat{X}_j, \hat{H}] \rangle + \langle[\hat{X}_{j} \hat{X}_i, \hat{H}] \rangle \right) \\
    &- \langle[\hat{X}_{i}, \hat{H}] \rangle \langle[\hat{X}_j, \hat{H}] \rangle \bigg) \\
	C_{ij} &= -i \bigg( \frac{1}{2} \left( \langle[\hat{X}_{i} \hat{X}_j, \hat{H}] \rangle + \langle[\hat{X}_{j} \hat{X}_i, \hat{H}] \rangle \right) \\
    &- \langle[\hat{X}_{i}, \hat{H}] \rangle \langle[\hat{X}_j, \hat{H}] \rangle \bigg)_{\bigg|_{V_{ij} =0}} 
\end{align*}
and the dissipative tensors,
\begin{align*}
	\frac{1}{2}\Gamma_{ij}^{kl} &= -\frac{d}{d {V}_{kl}} \bigg( \sum_n \gamma_n \frac{1}{2} \left( \langle \mathcal{D}_n[\hat{X}_i \hat{X}_j] \rangle + \langle \mathcal{D}_n[\hat{X}_j \hat{X}_i] \rangle\right)  \\ 
    &- \left(\langle \hat{X}_i \rangle \mathcal{D}_n[\hat{X}_j] +\langle \hat{X}_j \rangle \mathcal{D}_n[\hat{X}_i] \right)\bigg) \\
	\frac{1}{2}D_{ij} &= -\bigg( \sum_n \gamma_n \frac{1}{2} \left( \langle \mathcal{D}_n[\hat{X}_i \hat{X}_j] \rangle + \langle \mathcal{D}_n[\hat{X}_j \hat{X}_i] \rangle\right) \\
    &- \left(\langle \hat{X}_i \rangle \mathcal{D}_n[\hat{X}_j] +\langle \hat{X}_j \rangle \mathcal{D}_n[\hat{X}_i] \right)\bigg)_{\bigg|_{V_{ij} = 0}}
\end{align*}
We calculate now the different terms explicitly.

\begin{center}
\rule{0.5\linewidth}{0.4pt}
\end{center}
\emph{First-order terms} -- We start with the first-order terms, which are straightforward to calculate,
\begin{align*}
	&\langle [ \hat{X}_i, \hat{H}]\rangle = \sum_k {G}_{ik} \langle \hat{X}_k \rangle + C_i \\
	&\langle \hat{X}_{i}\rangle \langle[\hat{X}_j, \hat{H}] \rangle = \sum_{k} G_{jk} \langle \hat{X}_i\rangle  \langle \hat{X}_k\rangle + C_{j} \langle\hat{X}_i \rangle
\end{align*}

For the dissipators,
\begin{align*}
    	&\langle [ \hat{X}_i, \hat{H}]\rangle = \sum_k G_{ik} \langle \hat{X}_k \rangle + C_i \\
	&\langle \hat{X}_{i}\rangle \langle[\hat{X}_j, \hat{H}] \rangle = \sum_{k} G_{jk} \langle \hat{X}_i\rangle  \langle \hat{X}_k\rangle + C_{j} \langle\hat{X}_i \rangle
\end{align*}

\begin{center}
\rule{0.5\linewidth}{0.4pt}
\end{center}
\emph{Second-order terms} -- The second-order contributions are slightly more involved to calculate. 

For the Hamiltonian contributions, using the commutator identity $[\hat{A} \hat{B}, \hat{C}] = \hat{A}[\hat{B}, \hat{C}] + [\hat{A}, \hat{C}] \hat{B}$,
\begin{align*}
	\langle[\hat{X}_{i} \hat{X}_j, \hat{H}] \rangle = \langle\hat{X}_i \underbrace{[\hat{X}_j, \hat{H}]}_{\sum_k G_{jk} \hat{X}_k}\rangle + \langle\underbrace{[\hat{X}_i,\hat{H}]}_{\sum_l G_{ik} \hat{X}_k}\hat{X}_j\rangle \\
    = \sum_k G_{jk} \langle \hat{X}_i \hat{X}_j \rangle + \sum_l G_{ik} \langle \hat{X}_k \hat{X}_j \rangle
\end{align*}
Plugging it in,
\begin{align*}
	&G_{ij}^{kl} = \frac{d}{d {V}_{kl}} \sum_n \left[G_{in} \underbrace{\left(\frac{1}{2} \left (\langle \hat{X}_n \hat{X}_j \rangle + \langle \hat{X}_j \hat{X}_n \rangle \right) - \langle\hat{X}_j \rangle\langle\hat{X}_k\rangle \right)}_{= V_{jn}} \right. \\
	&+ \left. G_{jn} \underbrace{\left(\frac{1}{2} \left (\langle \hat{X}_n \hat{X}_i \rangle + \langle \hat{X}_i \hat{X}_n \rangle \right) - \langle\hat{X}_i \rangle\langle\hat{X}_k\rangle \right)}_{= V_{in}} \right] \\
    &= \frac{d}{dV_{kl}} \sum_n \left[ G_{in} V_{jn} + G_{jn} V_{in}  \right] 
\end{align*}

For the dissipators, calculating explicitly,
\begin{align*}
	&\langle \mathcal{D}_n[\hat{X}_i \hat{X}_j] \rangle = -\frac{1}{2}\delta_{n i} \langle\hat{X}_i \hat{X}_j\rangle \quad \text{for} \ i\neq j \\
	&\langle \mathcal{D}_n [\hat{X}_i^2]\rangle = -\delta_{ni} \left(\langle \hat{X}_i^2\rangle + \frac{1}{2} \right)
\end{align*}

Plugging it in,
\begin{align*}
  \frac{1}{2}\Gamma_{ij}^{kl} &= -\frac{d}{d {V}_{kl}} \sum_n \Bigg( 
    \gamma_n \frac{1}{2} \left( 
      \langle \mathcal{D}_n[\hat{X}_i \hat{X}_j] \rangle 
      + \langle \mathcal{D}_n[\hat{X}_j \hat{X}_i] \rangle 
    \right) \notag \\
  &\quad - \left( 
    \langle \hat{X}_i \rangle \mathcal{D}_n[\hat{X}_j] 
    + \langle \hat{X}_j \rangle \mathcal{D}_n[\hat{X}_i] 
  \right) \Bigg) \\ 
  &= \frac{1}{2} \frac{d}{d V_{kl}} \Bigg[ 
    \gamma_i \underbrace{ 
      \left( 
        \frac{1}{2} \left( 
          \langle \hat{X}_i \hat{X}_j \rangle 
          + \langle \hat{X}_j \hat{X}_i \rangle 
        \right) 
        - \langle \hat{X}_j \rangle \langle \hat{X}_i \rangle 
      \right) 
    }_{V_{ij}} \notag \\
  &\quad + \gamma_j \underbrace{ 
      \left( 
        \frac{1}{2} \left( 
          \langle \hat{X}_j \hat{X}_i \rangle 
          + \langle \hat{X}_j \hat{X}_i \rangle 
        \right) 
        - \langle \hat{X}_j \rangle \langle \hat{X}_i \rangle 
      \right) 
    }_{V_{ji}} 
  \Bigg] \\
  &= \frac{1}{2} \frac{d}{dV_{kl}} 
    \left( \gamma_i V_{ij} + \gamma_j V_{ji} \right) \\
  \frac{1}{2} D_{ii} &=  \gamma_i \left( V_{ii} - \frac{1}{2} \right) 
    \bigg|_{V_{ij} = 0} 
    = -\frac{1}{2}\gamma_i
\end{align*}

\begin{center}
\rule{0.5\linewidth}{0.4pt}
\end{center}
Finally, for the Hamiltonian terms,
\begin{equation}
    	G_{ij}^{kl} = \frac{1}{2} \left(\delta_{jl} G_{ki} + \delta_{kj} G_{li} + \delta_{li} G_{kj} + \delta_{ki}G_{lj} \right)
\end{equation}
For the dissipative terms,
\begin{align*}
	&\Gamma_{ij}^{kl} = -\frac{1}{2}(\gamma_i + \gamma_j)(\delta_{il}\delta_{jk} + \delta_{il}\delta_{jk}) \\
	&\Gamma_{ii}^{kl} = -\gamma_i\delta_{il}\delta_{ik} \\
	&D_{ij} = \delta_{ij} \gamma_i
\end{align*}

Putting it all together,
\begin{equation*}
\frac{d}{dt} {V}_{ij} = \sum_{kl} \left(G_{ij}^{kl} - \frac{1}{2} \Gamma_{ij}^{kl} \right){V}_{kl} + \left( C_{ij} - \frac{1}{2} D_{ij} \right)    
\end{equation*}

It is more convenient to write the equation in terms of matrix differential equation. We can write the Hamiltonian contributions in terms of the matrix product,
\begin{align}
	\sum_{kl} G_{ij}^{kl} V_{kl} &= \sum_l \left[ \underbrace{G_{il} V_{lj} }_{[\mathbf{G}\mathbf{V}]_{ij}} + \underbrace{G_{jl}V_{li}}_{[\mathbf{V}^T \mathbf{G}^T]_{ij}}] \right] = [\mathbf{GV} - (\mathbf{VG})^{\intercal}]_{ij} 
\end{align}
where we used the symmetries $\mathbf{V}^T = \mathbf{V}$ and $\mathbf{G}^T = -\mathbf{\Omega G \Omega}$.  

For the dissipative contributions,
\begin{align*}
  \sum_{kl} \Gamma_{ij}^{kl} V_{kl} - D_{ij} 
  &= (\gamma_i + \gamma_j) 
    \sum_{kl} \left( 
      \delta_{il}\delta_{jk} + \delta_{ik}\delta_{jl} 
    \right) V_{kl} 
    - \delta_{ij}\gamma_i \\
  &= (\gamma_i + \gamma_j)(V_{ij} + V_{ji}) 
    - \delta_{ij} \gamma_i \\
  &= 2 \underbrace{\sum_k \gamma_i \delta_{ik} V_{kj}}_{[\mathbf{\Gamma} \mathbf{V}]_{ij}} 
    + \underbrace{2 \sum_k \gamma_j \delta_{jk} V_{ki}}_{
      \substack{[\mathbf{\Gamma} \mathbf{V}]_{ji} = [(\mathbf{\Gamma} \mathbf{V})^T]_{ij} \\ = [\mathbf{V} \mathbf{\Gamma}]_{ij}}
    }
    - \underbrace{\delta_{ij} \gamma_i}_{[\mathbf{\Gamma}]_{ij}}
\end{align*}
\begin{center}
\rule{0.5\linewidth}{0.4pt}
\end{center}
Finally, putting it all together,
\begin{align}
	\frac{d}{dt} \mathbf{V} = [\mathbf{GV} + (\mathbf{VG})^\intercal] + \frac{1}{2}\left[\mathbf{\Gamma} - \left(\mathbf{\Gamma} \mathbf{V} + \mathbf{V} \mathbf{\Gamma} \right) \right]
\end{align}
here $[\mathbf{\Gamma}]_i = \gamma_i$ is a diagonal matrix of the resonator decay rates, This form emphasizes the similarity to the Lindblad equation. 

Another convenient form,
\begin{align}
	\frac{d}{dt} \mathbf{V} = \left[\left(\mathbf{\Omega H}- \frac{\mathbf{\Gamma}}{2} \right)\mathbf{V} + \mathbf{V}\left(\mathbf{\Omega H} -\frac{\Gamma}{2} \right)^\intercal \right] + \frac{\mathbf{\Gamma}}{2} 
\end{align}
where plugged in $\mathbf{G} = \mathbf{\Omega H}$. This is the form used in the main text.

Note that when we have only decay, this has a fixed point for the vacuum state $V_{ii} = 1/2$.
\begin{equation}
  	\frac{d}{dt} \text{tr} \mathbf{V} =  - \text{tr}(\mathbf{\Gamma}\mathbf{V}) + \frac{1}{2}\text{tr}(\mathbf{\Gamma}) = \sum_i \frac{\gamma_i}{2}(1 - 2V_{ii}).   
\end{equation}

\section{Solution and propagation matrices for common processes}

The solution for the equations can be formally written as the matrix exponential,
\begin{equation}
    \mathbf{S}(t) = \exp \left\{\left( \mathbf{\Omega H - \frac{\Gamma}{2}}\right) t\right\}
\end{equation} 
Unlike the simple symplectic case, the equations are no longer homogeneous, due to the free term $\mathbf{\Gamma/2}$ arising from the dissipative terms.

The first contribution to the solution is a homogeneous one obtained by propagating the mean vector and covariance matrix, 
\begin{align}
    \vec{\mu}_H (t) &= \mathbf{S}(t) \vec{\mu}_0 \\
    \mathbf{V}_H(t) &= \mathbf{S}(t) \mathbf{V}_0 \mathbf{S}^\intercal(t)
\end{align}
this contribution is similar to a simple symplectic evolution, just with added dissipation. It is analogous to the description of open quantum systems in terms of non-Hermitian Hamiltonians.

An additional contribution is due to the non-homogeneous term, physically reflecting the evolution of "vacuum fluctuations" leaking into the system from the environment,
\begin{align}
    &\vec{\mu}_\Gamma(t) = \int_0^t d\tau \mathbf{S}(t-\tau) \vec{C}(\tau) \\
    &\mathbf{V}_\Gamma(t) = \int_0^t d\tau \mathbf{S}(\tau) \left(\frac{\mathbf{\Gamma}}{2}\right)\mathbf{S}^\intercal(\tau)
\end{align}

\subsection{Squeezing under detuning}
In the IQ formulation, phase-sensitive amplification is described by the generators,
\begin{equation}
    \mathbf{\Omega H_{1}} = 
        \frac{r_a}{2} \begin{pmatrix}
            \sin\phi_a & -\cos\phi_a \pm \frac{\chi}{r_a} \\
            -\cos\phi_a \mp \frac{\chi}{r_a} & -\sin\phi_a
        \end{pmatrix}
\end{equation}

We can solve for the propagation matrix, giving us the IQ state generated by the interaction,
\begin{align}
    &\mathbf{S}_1(r,\phi; t)
    = e^{-\gamma_a t} \\
    &\begin{pmatrix}
        \cosh r_\chi t + \frac{r}{r_\chi}\sin\phi_a \sinh r_at & \frac{1}{r_\chi}(\chi - r_a \cos\phi_a)\sinh r_\chi t \\
        \frac{1}{r_\chi}(\chi + r_a\cos\phi_a)\sinh r_\chi t & \cosh r_\chi t - \frac{r}{r_\chi}\sin\phi_a \sinh r_\chi t
    \end{pmatrix}
    \notag
\end{align}
where $r_\chi \equiv \sqrt{r^2 -\chi^2}$. Note that the propagation matrix is almost symplectic, except for the decay factor. 

Note that the device is engineered to operate in the regime where the readout resonator is long-lived and $\gamma_a \ll r_a$, allowing us to treat the evolution as almost unitary over the time-scales of interest and neglect the decay factor. 

We can find the squeezing values and axis by calculating the covariance matrix of the IQ state and diagonalizing it. 
\begin{subequations}
\begin{align}
	&[\mathbf{V}_H]_{1,1} = \frac{e^{-\gamma_a t}}{2 r_\chi^2} \left[ r^2 \cosh 2r_\chi t  + r r_\chi \sinh 2r_\chi t -\chi^2 \right] \\
	&[\mathbf{V}_H]_{2,2} = \frac{e^{-\gamma_a t}}{2 r_\chi^2} \left[ r^2 \cosh 2r_\chi t  - r r_\chi \sinh 2r_\chi t - \chi^2 \right] \\
	&[\mathbf{V}_H]_{1,2} = \frac{e^{-\gamma_a t}}{2 r_\chi^2} \left[  2\chi r \sinh^2 r_\chi t \right]
\end{align}
\end{subequations}
due to the detuning the covariance matrix is not diagonal, and the squeezing is not along the axis defined by $\phi_a$. To find the effective squeezing angle, we can diagonalize the covariance matrix and find its eigenvectors, which allows us to calculate the squeezing angle relative to the $I$-axis,
\begin{equation}
    	\tan \phi_\chi = \frac{1}{\chi \sinh r_\chi t} \left(r_\chi \cosh r_\chi t +  \sqrt{r^2 \cosh^2 r_\chi t - \chi^2}  \right).
\end{equation}
with squeezing values,
\begin{align*}
	\lambda_{\pm} &= \frac{1}{2r_\chi^2} \bigg[ (r^2\cosh 2r_\chi t - \chi^2)  \\
    &\pm  r \sinh r_\chi t \sqrt{2\left(r^2 \cosh^2  r_\chi t - 2 \chi ^2\right)}   \bigg]
\end{align*}

Generally, since the system is open, the covariance matrix would have an additional contribution arising from the vacuum fluctuations leaking into the resonator. This one is negligible in our regime of operation, but we write here for completeness.
\begin{align}
	&V_P^{(1,1)} = \frac{\gamma_a}{r_\chi^3} \left(r^2 \sinh 2 r_\chi t + 2r r_\chi \sinh ^2r_\chi t -2 \chi ^2 r_\chi t \right)  \\
	&V_P^{(2,2)} = \frac{\gamma_a}{4 r_\chi^3}  \left(r^2 \sinh 2 r_\chi t - 2r r_\chi \sinh^2 r_\chi t- 2 \chi ^2 r_\chi t \right) \\
	&V_P^{(1,2)} =\frac{\gamma_a r \chi}{4 r_\chi^3}(\sinh 2 r_\chi t-2 r_\chi t)
\end{align}

\subsection{Conversion under detuning}
This conversion can be similarly described in IQ-space for a full solution. In the IQ formulation, we need to now work in a 2-mode basis $(\vec{X}_a, \vec{X}_s) = (I_a, Q_a, I_s, Q_s)$. 
\begin{equation}
    \mathbf{\Omega H_{3}} = 
\begin{pmatrix} 
	0 & \mp\chi & \sin \phi_3 & \cos \phi_3 \\
	\pm\chi & 0 & -\cos \phi_3 & \sin \phi_3 \\
	-\sin \phi_3 & \cos \phi_3 & 0 & 0 \\
	-\cos\phi_3 & -\sin\phi_3 & 0 & 0
\end{pmatrix}
\end{equation}
The propagation matrix generated by this admits an analytic expression, but it is rather complex and not informative.

\section{Pulse shapes}
For the numerical calculations in this manuscript we assumed Gaussian pulses of the form,
\begin{equation}
    g_{a, \phi, t_1, t_2, \nu_1, \nu_2}(t) = \frac{ae^{-i\phi}}{4} 
\operatorname{erfc}\left( -v_1 (t - t_1) \right) 
\operatorname{erfc}\left( v_2 (t - t_2) \right)
\end{equation}
where $\nu_{1(2)}$ are the rise/fall rates of the pulse, $a$ is the amplitude of the pulse, and $t_{1(2)}$ are the start and end times of the pulse.

\begin{figure}
    \centering
    \includegraphics[width=0.85\linewidth]{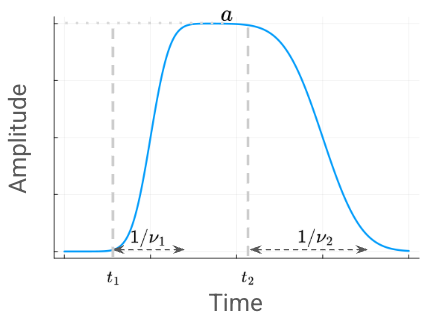}
    \caption{A general pulse with the different parameters illustrated on the figure.}
    \label{fig:pulse_shape}
\end{figure}

This allows smooth parametrization of the pulses, while maintaining smooth rise-up and rise-down pulse phases.

\section{Angular-momentum representation of conversion}
We write the conversion Hamiltonian in terms of generalized angular momentum operators using a Schwinger mapping \cite{schwinger_angular_1952},
\begin{equation}
    \hat{H}_{conv} = \pm\frac{\chi}{2} \hat{N} + g_{as}\cos \theta_{as} \hat{J}_x - g_{as} \sin\theta_{as}\hat{J}_y \pm  \chi\hat{J}_z
\end{equation}
The angular momentum operators define the relative phase coherence between the two oscillators ($\hat{J}_x$ and $\hat{J}_y$), the difference in the number of photons between the two oscillators ($\hat{J}_z$). They are defined as,
\begin{subequations}
    \begin{align}
        \hat{J}_x &= \frac{1}{2} \left ( \hat{a}^\dagger \hat{s} + \hat{s}^\dagger \hat{a} \right ) \\
        \hat{J}_y &= -\frac{i}{2} \left ( \hat{a}^\dagger \hat{s} - \hat{s}^\dagger \hat{a} \right ) \\
        \hat{J}_z &= \frac{1}{2} \left ( \hat{a}^\dagger \hat{a} - \hat{s}^\dagger \hat{s} \right ) \\
        \hat{J}^2 &= \hat{J}_x^2 + \hat{J}_y^2 + \hat{J}_z^2 = \frac{\hat{N}}{2} \left [ \frac{\hat{N}}{2} + 1 \right ]
    \end{align}
\end{subequations}
where $\hat{N} = \hat{a}_1^\dagger \hat{a}_1 + \hat{a}_2^\dagger \hat{a}_2$ is the total number of photons in the two resonators, and $\hat{J}^2$ is analogous to the magnitude of the angular momentum, and reflects the conversion of the total number of photons by the conversion process.

Up to an overall phase, the conversion process behaves like a generalized spin, precessing around a generalized "Bloch" vector $\vec{\Omega} = (g_{as}\cos \theta_{as}, -g_{as} \sin \theta_{as}, \pm\chi)$.

\section{Optimization and readout metrics}
\label{app:optimization}
\subsection{Linear Discriminant Analysis}
Linear discriminant analysis (LDA), also known as the Fisher Discriminant method, is a dimensionality reduction technique used to find a projection plane which maximizes the separation between a set of Gaussian distributions. We will focus here specifically on the case of two distributions, which we interpret as the two possible readout distributions, associated with the ground and excited states.

The objective function for LDA  is the Fisher Index, defined as the separation of the projected means divided by the sum of the scattering of the two classes.
\begin{equation}
        J(\vec{w}) = \frac{\vec{w}^T (\vec{\mu}_1 - \vec{\mu}_2) (\vec{\mu}_1 - \vec{\mu}_2)^T \vec{w}}{\vec{w}^T(\mathbf{V}_1 + \mathbf{V}_2) \vec{w}} 
\end{equation}
where here the indices $1,2$ refer to the excited and ground states. 

We define the between-class scatter $\mathbf{S}_b = (\vec{\mu}_1 - \vec{\mu}_2) (\vec{\mu}_1 - \vec{\mu}_2)^T$ and the in-class scatter $\mathbf{S_w} = \mathbf{V}_1 + \mathbf{V}_2$, which allows us to write,
\begin{equation}
         J(\mathbf{w}) = \frac{\mathbf{w}^T \mathbf{S}_b \mathbf{w}}{\mathbf{w}^T \mathbf{S}_w \mathbf{w}}
\end{equation}

We optimize this with respect to $\mathbf{w}$, taking the derivative $\partial_{\mathbf{w}}$ and equating to 0. This gives us the following (generalized) eigenvalue problem,
\begin{equation}
    \mathbf{S}_b \vec{w} = J(\vec{w}) \mathbf{S}_w \vec{w}
\end{equation}
which indicates that the maximum value would the Fisher Index is the highest eigenvalue of the problem, and the associated eigenvector is the optimal separating plane. Solving this, we find that the projection axis $\vec{w}_o = \frac{\vec{\mu}_1 - \vec{\mu}_0}{\sigma_1 + \sigma_0}$ maximizes the Fisher Index when it is parallel to the mean separation, normalized by the sum of projected variances. This defines the Fisher Discriminant, which is the optimal value of the Fisher Index.
\begin{equation}
    \mathcal{J}(\vec{\theta}) = \frac{\Delta\vec{\mu}^\intercal \Delta\vec{\mu}}{\Delta\vec{\mu}^\intercal ( \mathbf{V}_1 + \mathbf{V}_2 ) \Delta\vec{\mu}} = \frac{(m_1(\vec{w}) - m_2(\vec{w}))^2}{\sigma^2_1(\vec w) + \sigma^2_2(\vec{w})}
\end{equation}
where $\Delta \vec{\mu} = \vec{\mu}_1 - \vec{\mu}_2$ and the different system parameters are grouped formally in $\vec\theta$.  We denote by $m_{1(2)}$ and $\sigma_{1(2)}$ the means and variances projected on the line separating the two distributions. 

This gives a simple interpretation for the Fisher Discriminant - it is simple the mean separation, divided by the noise projected over the separation axis.

\subsection{Fidelity}
Fidelity is a common metric in quantum computing for quantifying the error of a components. It is defined as $F = 1 - P_{err}$, where $P_{err}$ is the probability of making an error. Specifically in readout, it is the probability of misclassifying a state.

In the multivariate case, the integrals and threshold can be difficult to calculate, but for Gaussian states the definition can be used in conjunction with LDA to simplify the calculation. Using LDA, the problem can be reduced to a one-dimensional problem, describing the distributions projected along the separation axis,
\begin{align}
	P_{1(2)}(x) = \frac{1}{\sqrt{2\pi \sigma_{1(2)}^2}} \exp \left(- \frac{(x-\mu_{1(2)})^2}{2 \sigma_{1(2)}^2} \right)
\end{align}

Readout is essentially distinguishing between two states, where we obtain a value along and try to identify the most likely distribution to have produced it. We can define a threshold value, for which both distributions are equally likely, 
\begin{equation}
    	P_1(x_{th}) = P_2(x_{th})
\end{equation}
this gives a quadratic equation for the threshold value,
\begin{equation}
    	\frac{(x_{th} - \mu_1)^2}{2 \sigma_1^2} - \frac{(x_{th} - \mu_2)^2}{2 \sigma_2^2} = \frac{1}{2} \ln \left( \frac{\sigma_2^2}{\sigma_1^2}\right).
\end{equation}

This defines a threshold value for distinguishing between the two state -- values above the threshold would more likely belong to one distribution, while values below the threshold would more likely belong to another distribution.

The error probability is the given by the overlaps of the two distribution, from the equal likelihood point.
\begin{equation}
    	P_{err} = \frac{1}{2} \left[ \text{erfc}\left(\frac{x_{th} - \mu_1}{\sqrt{2}\sigma_1} \right) + \text{erfc}\left(\frac{ \mu_2 - x_{th}}{\sqrt{2}\sigma_2} \right) \right] 
\end{equation}

A particularly relevant case is when the projected variances are equal and $\sigma_0 = \sigma_1 = \sigma$. In this case, by symmetry $x_{th} = \frac{1}{2} (\mu_0 + \mu_1)$. 

Plugging this into the expression for the fidelity,
\begin{align}
	&P_{err} =  \text{erfc} \left( \frac{\mu_2 - \mu_1}{2 \sqrt{2} \sigma} \right) = \text{erfc} \left( \frac{1}{2} \sqrt{\frac{\mathcal{J}}{2}} \right) \\
	&F = \text{erf}\left(\frac{1}{2} \sqrt{\frac{\mathcal{J}}{2}} \right)
\end{align}

\subsection{Formal objective function}

Based on these two metrics, we define the principle objective function as the product of the two:
\begin{equation}
    \mathcal{L}_0(\vec{\theta}) = \lambda_1\mathcal{J}_o(\vec{\theta}) + \lambda_2\mathcal{G}_{ro}(\vec{\theta}).
\end{equation}
where $\lambda_{1,2}$ are weight parameters weighing the relative contribution of the separability and gain to the objective function.

Formally, we can incorporate this into the objective function by a correction factor:
\begin{equation}
    \mathcal{L}\left(\vec{\theta}\right) = \left(1 - \frac{\max_t n_a}{n_a^{\rm crit}}  - \sum_{\omega_i \omega_j}  \frac{\max_t g_{\omega_i \omega_j}}{g^{\rm crit}_{\omega_i \omega_j}} \right) \mathcal{L}_0(\vec{\theta})
\end{equation}
This product form ensures that readout backaction cannot be compensated by good readout parameters.

This function guides the design of the SNAIL-resonator system, ensuring that the system is tuned for optimal performance. We use this objective function to analyze our different readout schemes using numerical simulations and theoretical analysis, and relevant operation regimes.

\section{Multiplexing and Frequency crowding}
In frequency-multiplexed superconducting circuits with nonlinear interactions (e.g., driven by a SNAIL), coherent interactions are selectively activated by applying drive tones at the sum or difference of oscillator frequencies. The goal is to ensure each interaction is uniquely addressable, without accidentally activating other unwanted modes.

Let $\omega_1, \omega_2, \ldots, \omega_N$ be the resonator frequencies, with $N$ being the number of resonators in the system.

Each interaction, in the linear approximation, is triggered by drive tones at sum and difference frequencies:
\begin{equation}
    \omega_{drive} = \omega_i \pm \omega_j, \qquad i\neq j
\end{equation}

We want each drive to only drive a single unique interaction. This determines two constraints:
\begin{enumerate}
    \item No drive tone coincides with an oscillator frequency -- $\omega_k \neq \omega_i \pm \omega_j \quad \forall \ i\neq j$.
    \item Each interaction is unique -- $\omega_i \pm \omega_j \neq \omega_k \pm \omega_l$.
\end{enumerate}

We denote the set of resonator frequencies by $\Omega$,
\begin{equation}
    \Omega = \{\omega_1, \ldots, \omega_N\}
\end{equation}
where we recall that $N$ is the number of resonators in the system. 

Similarly, we denote by $\mathcal{M}$ the set of different mixing frequencies,
\begin{equation}
    \mathcal{M} = \{\omega_i + \omega_j, \omega_i - \omega_j | i \neq j \}
\end{equation}

Our constraints, formalized mathematically, imply that $\Omega \cap \mathcal{M} = \emptyset$. Additionally, $\mathcal{M}$ should include all the different sums and difference frequencies. Simple counting arguments show that the number of mixing frequencies is,
\begin{equation}
    |\mathcal{M}| = N(N-1)
\end{equation}

To keep everything uniquely addressable, we need to fit the frequencies in the set $\Omega$ and $\mathcal{M}$ into this range,
\begin{equation}
    |\mathcal{M}| + |\Omega| = N(N-1) + N = N^2.
\end{equation}

In other words, we need to fit $N^2$ different frequency bands into the bandwidth,
\begin{equation}
    \text{BW} > N^2 \cdot \delta\omega
\end{equation}
where $\text{BW}$ is the bandwidth required for uniquely addressable interactions and resonators. This scaling gets even worse if we consider higher-order interactions, and the scaling would now go as $N^k$, where $k$ is the order of the interactions.

Finding a set of frequencies that satisfies these constraints is a classic constraint studied in additive number theory, where special sets such as Sidon sets (a set where all pairwise sums are distinct) and Golomb rulers (a set where all pairwise differences are distinct) are used to eliminate these kinds of collisions. These constructions guarantee that the drive tones do not overlap accidentally.

\section{Derivation of the squared-SNR and noise kernel}
As in the main text, we define the Signal-to-Noise Ratio (SNR), denoted as $D^2$, for the discrimination between two qubit states (ground $|g\rangle$ and excited $|e\rangle$) using the integrated homodyne current:

\begin{equation}
    \mathcal{D}^2 = \frac{(\langle \hat{z} \rangle_e - \langle \hat{z} \rangle_g)^2}{\text{Var}[\hat{z}]}   
\end{equation}

where the integrated current operator $\hat{z}$ is defined with a filter function $g(t)$ over the integration time $T$:

\begin{equation}
    \hat{z} = \int_0^T g(t)\hat{I}(t) dt    
\end{equation}

The numerator of the SNR depends on the difference in the mean integrated currents for the two qubit states. Let $\Delta I(t) = \langle \hat{I}(t) \rangle_e - \langle \hat{I}(t) \rangle_g$ be the signal contrast. The signal difference is:

\begin{equation}
    \Delta z = \langle \hat{z} \rangle_e - \langle \hat{z} \rangle_g = \int_0^T g(t) \Delta I(t) dt    
\end{equation}

If the optimal filter is chosen such that $g(t) \propto \Delta I(t)$, this integral maximizes the signal-to-ratio in the linear case for Gaussian white noise.

The variance of the integrated signal is given by the double integral of the correlation function:
\begin{equation}
    \text{Var}[\hat{z}] = \langle \hat{z}^2 \rangle - \langle \hat{z} \rangle^2 = \int_0^T dt_1 \int_0^T dt_2 \, g(t_1)g(t_2) \Gamma_{II}(t_1, t_2)   
\end{equation}
where $\Gamma_{II}(t_1, t_2) = \langle \hat{I}(t_1)\hat{I}(t_2) \rangle - \langle \hat{I}(t_1) \rangle \langle \hat{I}(t_2) \rangle$ is the current-current correlation function.

Using the input-output relation for the homodyne current $\hat{I}(t) = \sqrt{2\eta\gamma}\hat{X}(t) + \hat{\xi}(t)$ (where $\hat{\xi}(t)$ represents the white noise background) , we decompose the correlation function into a system part and a background noise part:

\begin{equation}
    \Gamma_{II}(t_1, t_2) = 2\eta\gamma \Gamma_{XX}(t_1, t_2) + \delta(t_1 - t_2)\left( \frac{1}{2} + N_{\text{add}} \right)    
\end{equation}
Here, $\Gamma_{XX}(t_1, t_2)$ is the intracavity noise kernel derived in the previous section, and the delta function represents the shot noise limit (plus any added technical noise $N_{\text{add}}$).

Substituting this back into the variance integral:
\begin{equation}
\begin{split}
    \text{Var}[\hat{z}] &= \left( \frac{1}{2} + N_{\text{add}} \right) \int_0^T g^2(t) dt \\
    &\quad + 2\eta\gamma \int_0^T dt_1 \int_0^T dt_2 \, g(t_1)g(t_2) \Gamma_{XX}(t_1, t_2)
\end{split}
\label{eq:variance_split}
\end{equation}
This expression explicitly relates the readout fidelity to the intracavity noise kernel $\Gamma_{XX}$, capturing the effects of finite bandwidth and squeezing transients.

\subsection{Derivation of the noise kernel}
In order to derive an expression for the noise kernel, we consider a passive cavity with energy decay rate $\gamma$ and detuning $\Delta$ from the drive frequency. The Heisenberg equation of motion for the intracavity annihilation operator $\hat{a}$ in the rotating frame is given by the Quantum Langevin equation:
\begin{equation}
    \frac{d\hat{a}}{dt} = -\left(\frac{\gamma}{2} + i\Delta\right)\hat{a}(t) + \sqrt{\gamma}\hat{a}_{\text{in}}(t)  
\end{equation}
where $\hat{a}_{\text{in}}(t)$ is the input noise operator satisfying the vacuum commutation relations $[\hat{a}_{\text{in}}(t), \hat{a}_{\text{in}}^\dagger(t')] = \delta(t-t')$.

Defining the complex damping parameter $\Lambda = \frac{\gamma}{2} + i\Delta$, the formal solution for the fluctuation operator $\Delta\hat{a}(t) = \hat{a}(t) - \langle\hat{a}(t)\rangle$ starting from time $t=0$ is:
\begin{equation}
    \Delta\hat{a}(t) = \Delta\hat{a}(0)e^{-\Lambda t} + \sqrt{\gamma}\int_0^t e^{-\Lambda(t-\tau)}\hat{a}_{\text{in}}(\tau) d\tau  
\end{equation}

We aim to calculate the two-time correlation function (noise kernel) for the intracavity quadrature $\hat{X} = \frac{1}{\sqrt{2}}(\hat{a} + \hat{a}^\dagger)$. We define the kernel as the symmetric covariance:
\begin{equation}
    \Gamma_{XX}(t_1, t_2) = \frac{1}{2}\langle \{\Delta\hat{X}(t_1), \Delta\hat{X}(t_2)\} \rangle  
\end{equation}
However, for standard ordered calculations, it is convenient to expand this using the annihilation and creation operators. Assuming $t_1 \ge t_2$, we decompose the kernel into \emph{normal} and \emph{anomalous} ordering contributions:
\begin{equation}
    \Gamma_{XX}(t_1, t_2) = \text{Re}\left[ \Gamma_{aa^\dagger}(t_1, t_2) \right] + \text{Re}\left[ \Gamma_{aa}(t_1, t_2) \right]    
\end{equation}
where we have defined the correlation components:
\begin{subequations}
\begin{equation}
    \Gamma_{aa^\dagger}(t_1, t_2) = \langle \Delta\hat{a}(t_1)\Delta\hat{a}^\dagger(t_2) \rangle
\end{equation}
\begin{equation}
    \Gamma_{aa}(t_1, t_2) = \langle \Delta\hat{a}(t_1)\Delta\hat{a}(t_2) \rangle
\end{equation}  
\end{subequations}

Substituting the formal solution into the definition of $\Gamma_{aa^\dagger}$:
\begin{equation}
\begin{split}
    &\Gamma_{aa^\dagger}(t_1, t_2) = \langle \Delta\hat{a}(0)\Delta\hat{a}^\dagger(0) \rangle e^{-\Lambda t_1} e^{-\Lambda^* t_2} \\
    &+ \gamma \int_0^{t_1} d\tau \int_0^{t_2} d\tau' e^{-\Lambda(t_1-\tau)} e^{-\Lambda^*(t_2-\tau')} \langle \hat{a}_{\text{in}}(\tau)\hat{a}_{\text{in}}^\dagger(\tau') \rangle
\end{split}
\label{eq:gamma_aadag_split}
\end{equation}

For vacuum input noise, the correlator is $\langle \hat{a}_{\text{in}}(\tau)\hat{a}_{\text{in}}^\dagger(\tau') \rangle = \delta(\tau-\tau')$. The double integral simplifies to a single integral over the interval $[0, t_2]$.

Using $\Lambda + \Lambda^* = \gamma$, the integral yields $\frac{1}{\gamma}(e^{\gamma t_2} - 1)$. Thus, the noise contribution is:
\begin{equation}
    \mathcal{I}_{\text{noise}} = e^{-\Lambda(t_1-t_2)} (1 - e^{-\gamma t_2})   
\end{equation}

Combining this with the initial state decay term, and letting $\tau = t_1 - t_2$:
\begin{equation}
    \Gamma_{aa^\dagger}(t_1, t_2) = e^{-\Lambda \tau} \left[ 1 + e^{-\gamma t_2}(\langle \Delta\hat{a}\Delta\hat{a}^\dagger \rangle_0 - 1) \right]    
\end{equation}

Substituting the formal solution into $\Gamma_{aa}$:
\begin{equation}
    \Gamma_{aa}(t_1, t_2) = \langle \Delta\hat{a}^2 \rangle_0 e^{-\Lambda(t_1+t_2)}
\end{equation}
Since for a vacuum or thermal bath, the anomalous noise correlator vanishes: $\langle \hat{a}_{\text{in}}(\tau)\hat{a}_{\text{in}}(\tau') \rangle = 0$. Therefore, the anomalous term is purely determined by the decay of the initial squeezing.

We can relate  the initial operator expectations to the standard quadrature variances $V_{XX}, V_{PP}$ and covariance $C_{XP} = \frac{1}{2}\langle \{\Delta\hat{X}, \Delta\hat{P}\} \rangle$ at $t=0$. Using $\Delta\hat{a} = \frac{1}{\sqrt{2}}(\Delta\hat{X} + i\Delta\hat{P})$:
\begin{equation}
    \langle \Delta\hat{a}\Delta\hat{a}^\dagger \rangle_0 = \frac{1}{2}(V_{XX} + V_{PP} + 1)  
\end{equation}

\begin{equation}
    \langle \Delta\hat{a}^2 \rangle_0 = \frac{1}{2}(V_{XX} - V_{PP} + 2iC_{XP})    
\end{equation}
where we used $\langle \{\Delta\hat{X}, \Delta\hat{P}\} \rangle = 2C_{XP}$).

Substituting the initial conditions back into the kernel components and taking the real parts, we arrive at the final expression for the intracavity noise kernel.

\begin{equation}
    \text{Re}[\Gamma_{aa^\dagger}] = e^{-\frac{\gamma}{2}\tau}\cos(\Delta\tau) \left[ 1 + \frac{1}{2}e^{-\gamma t_2}(V_{XX} + V_{PP} - 1) \right]    
\end{equation}

\begin{equation}
\begin{split}
    \text{Re}[\Gamma_{aa}] &= \frac{1}{2}e^{-\frac{\gamma}{2}(t_1+t_2)} \Big[ (V_{XX}-V_{PP})\cos(\Delta(t_1+t_2)) \\
    &\quad + 2C_{XP}\sin(\Delta(t_1+t_2)) \Big]
\end{split}
\label{eq:gamma_aa_split}
\end{equation}

Summing these contributions yields the total kernel:
\begin{equation}
\begin{split}
    &\Gamma_{XX}(t_1, t_2) = \frac{1}{2} e^{-\frac{\gamma}{2}\tau}\cos(\Delta\tau) \left[ 2 + e^{-\gamma t_2}(V_{XX} + V_{PP} - 1) \right] \\
    &\quad + \frac{1}{2} e^{-\frac{\gamma}{2}(t_1+t_2)} \big[ (V_{XX}-V_{PP})\cos(\Delta(t_1+t_2)) \\
    &\quad + 2C_{XP}\sin(\Delta(t_1+t_2)) \big]
\end{split}
\label{eq:gamma_xx_final}
\end{equation}
where $\tau = t_1 - t_2 \ge 0$.

\end{document}